\documentclass{article}

\usepackage[utf8]{inputenc} %unicode support
\usepackage{dsfont} 
\usepackage{cite}
\usepackage{amsthm,amsmath,amsfonts,amssymb}
\usepackage[nohyperlinks]{acronym}
\usepackage{color}
\usepackage{nicefrac}
\usepackage{graphicx}
\DeclareGraphicsExtensions{.pdf,.jpeg,.png}
\graphicspath{{./graphics/}}
\usepackage{lipsum}
\usepackage{microtype}
\usepackage{tikz}
\usetikzlibrary{arrows}
\usetikzlibrary{math}
\usetikzlibrary{calc}
\tikzset{>=latex}
\usetikzlibrary{decorations.pathreplacing}
\usetikzlibrary{shapes}
\usetikzlibrary{shapes.callouts}
\usetikzlibrary{arrows.meta}
\usetikzlibrary{shapes,snakes}

\usepackage{float}

\newcommand{\Mat}[1]{\boldsymbol{#1}}
\newcommand{\be}{\begin{equation}}
\newcommand{\ee}{\end{equation}}

\DeclareMathOperator*{\argmax}{arg\,max}

\begin{document}

\title{Better than Rician: Modelling millimetre wave channels as Two-Wave with Diffuse Power}

\author{Erich Z\"ochmann\footnote{Christian Doppler Laboratory for Dependable Wireless Connectivity for the Society in Motion, Institute of Telecommunications, TU Wien}, Sebastian Caban, Christoph F. Mecklenbr\"auker, \\ Stefan Pratschner, Martin Lerch, Stefan Schwarz, Markus Rupp}

\maketitle

\begin{abstract} % abstract

This contribution provides experimental evidence for the two-wave with diffuse power {(TWDP)} fading model.
We have conducted two indoor millimetre wave measurement campaigns with directive horn antennas at both link ends. 
One horn antenna is mounted in a corner of our laboratory, while the other is steerable and scans azimuth and elevation.
Our first measurement campaign is based on scalar network analysis with $7$\,GHz of bandwidth.
Our second measurement campaign obtains magnitude and phase information, additionally sampled directionally at several positions in space.
%This second measurement campaign is limited to 2 GHz bandwidth.
We apply Akaike's information criterion to decide whether Rician fading sufficiently explains the data or the generalized {TWDP} fading model is necessary.
Our results indicate that the TWDP fading hypothesis is favoured over Rician fading in situations where the steerable antenna is pointing towards reflecting objects or is slightly misaligned at line-of-sight.
We demonstrate {TWDP} fading in several different domains, namely, frequency, space, and time.

\end{abstract}

\section{Introduction}

Accurate modelling of wireless propagation effects is a fundamental prerequisite for a proper communication system design.
After the introduction of the double-directional radio channel model~\cite{steinbauer2001double}, wireless propagation research ($<6$\,GHz) started to model the wireless channel agnostic to the antennas used. 
More than a decade later, propagation research focusses now on millimetre wave bands to unlock the large  bandwidths available in this regime~\cite{anderson2004building,moraitis2006measurements,geng2009millimeter,wangchuk2017double}. 
At \acp{mmWave}, omnidirectional antennas have small effective antenna areas, resulting in a high path-loss~\cite{peter2015path,peter2016channel,samimi2015probabilistic,deng201528,maccartney2015millimeter}. 
To overcome this high path-loss, researchers have proposed to apply highly directive antennas on both link ends~\cite{roh2014millimeter,hur2013millimeter,sun2014mimo,pi2011introduction}. 
Most researchers aim to achieve high directivity with antenna arrays~\cite{heath2016overview,andrews2017modeling,alkhateeb2014channel,kulkarni2016comparison,zochmann2016comparing,pratschner2017mutual}
and a few with dielectric lenses~\cite{brady2013beamspace,zeng2016millimeter,zeng2017cost}.
When the link-quality depends so much on the achieved beam-forming gain, antennas must be considered as part of the wireless channel again.  
Small-scale fading is then influenced by the antenna.

According to Durgin~\cite[p. 137]{durgin2003space}, ``The use of directive antennas or arrays at a receiver, for example, amplifies several of the strongest multipath waves that arrive in one particular direction while attenuating the remaining waves. This effectively increases the ratio of specular to nonspecular received power, turning a Rayleigh or Rician fading channel into a TWDP fading channel.''
The mentioned \ac{TWDP} fading channel describes this spatial filtering effect by two non-fluctuating receive signals together with many smaller diffuse components.

\subsection{Related work}

The authors of \cite{dupleich2017investigations} investigated a simple wall scattering scenario and analysed how fading scales with various antenna directivities and different bandwidths. 
Increasing directivity~\cite{dupleich2017investigations}, as well as increasing bandwidth~\cite{dupleich2017investigations,iqbal2017stochastic}, results in an increased Rician K-factor.
The authors of~\cite{samimi201628} analysed fading at $28$\,GHz with high gain horn antennas on both link ends. 
They observe high Rician K-factors even at \ac{NLOS}.  This effect is explained by spatial filtering of directive antennas, as they suppress many multipath components~\cite{dupleich2017investigations}.
Outdoor measurements in~\cite{sun2017millimeter,rappaport2017small}, show  a graphical agreement with the Rice fit, but especially Fig. 10 in \cite{rappaport2017small} might be better explained as \ac{TWDP} fading.

\ac{TWDP} fading has already successfully been applied to describe $60$\,GHz near body shadowing~\cite{mavridis2015near}. 
Furthermore, as quoted above, \ac{TWDP} must be considered for arrays, as they act as spatial filters~\cite{durgin2002new, durgin2003space}. 
%The authors of \cite{rappaport2014millimeter} argue for the use of \ac{TWDP} in \ac{mmWave} as well. 
While theoretical work on \ac{TWDP} fading is already advanced~\cite{esposito1973statistical,oh2005ber,saberali2013new,
rao2015mgf,romero2017fluctuating,schwarz2017outage,Schwarz_VTC17}, experimental evidence, especially at millimetre waves, is still limited.
For enclosed structures, such as aircraft cabins and buses, the applicability of the \ac{TWDP} model is demonstrated by Frolik~\cite{frolik2007case,frolik2008appropriate,frolik2009compact,matolak2011worse,bakir2009diversity}.
A deterministic two ray behaviour in ray tracing data of \ac{mmWave} train-to-infrastructure communications is shown in~\cite{Zoe2017TwoRay}.
A further extension of the \ac{TWDP}-fading model, the so called fluctuating two-ray fading model, was also successfully applied to fit \ac{mmWave} measurement data~\cite{romero2017fluctuating}.

Our group has conducted two  measurement campaigns \cite{Zoe2016Direct,Zoe2017Assoc} to directionally analyse receive power and small-scale fading parameters for \acp{mmWave}. 
This contribution is based on~\cite{Zoe2016Direct,Zoe2017Assoc}.

\subsection{Outline and contributions}

With this contribution, we aim to bring scientific rigour to the small-scale fading analysis of millimetre wave indoor channels. 
We show in Section~\ref{sec:meth}\,--\,by means of an information-theoretic approach~\cite{burnham2003model} and null hypothesis testing~\cite{frick1996appropriate}\,--\,that the \ac{TWDP} model has evidence in \ac{mmWave} communications. 

We have conducted two measurement campaigns within the same laboratory with different channel sounding concepts. 
Our measurements are carried out in the V-band; the applied center frequency is $60$\,GHz.
For both measurement campaigns, $20$\,dBi horn antennas are used at the transmitter and at the receiver.    
The \ac{MC1} samples the channel in azimuth ($\varphi$) and elevation ($\theta$), keeping the antenna's (apparent) phase center~\cite[pp.
799]{balanis2005antenna} at a fixed $(x,y)$\,--\,coordinate. 
The transmitter is mounted in a corner of our laboratory. 
The sounded environment as well as the mechanical set-ups are explained in Section~\ref{sec:mech}.
For \ac{MC1}, we sounded the channel in the frequency-domain by aid of scalar network analysis, described in Section~\ref{sec:scalar}. 
These channel measurements span over $7$\,GHz bandwidth, supporting us to analyse fading in the \emph{frequency domain}.  

For the \ac{MC2}, described in Section~\ref{sec:vec}, we improved the set-up mechanically and \ac{RF}\,--\,wise. 
By adding another linear guide along the $z$-axis, we keep the antenna's phase center constant in $(x,y,z)$\,--\,coordinate, irrespective of the antenna's elevation. 
Furthermore, we changed the sounding concept to time-domain channel sounding. 
This approach allows us to utilise the \emph{time domain} and to show channel impulse responses in Section~\ref{sec:CIR}. 
Additionally, by adjusting $(x,y,z,\varphi,\theta)$, we sample the channel in the \emph{spatial domain} at all directions $(\varphi, \theta)$.
These improvements enable us to show spatial correlations in Section~\ref{sec:spatial}, a further analysis tool to support the claims from \ac{MC1}.

In summary, we \emph{demonstrate \ac{TWDP} fading} for directional \ac{mmWave} indoor channels in the \emph{frequency-domain}, in the \emph{spatial-domain}, and in the \emph{time-domain}. 

\section{Methodology - Fading model identification}
\label{sec:meth}
\ac{TWDP} fading captures the effect of interference of two non-fluctuating radio signals and many smaller so called diffuse signals~\cite{durgin2002new}. 
The \ac{TWDP} distribution degenerates to Rice if one of the two non-fluctuating radio signals vanishes. 
This is analogous to the well known Rice degeneration to the Rayleigh distribution with decreasing K factor.
%and Rayleigh fading are limiting cases of this model. 
In the framework of model selection, \ac{TWDP} fading, Rician fading, and Rayleigh fading are hence nested hypotheses~\cite{burnham2003model}.
Therefore, it is also obvious that among these alternatives, \ac{TWDP} always allows the best possible fit of measurement data.
Occam's razor~\cite{berger1992application} asks to select, among competing hypothetical distributions, the hypothesis that makes the fewest assumptions.
Different distribution functions are often compared via a goodness-of-fit test~\cite{maydeu2010goodness}. 
Nevertheless, the authors of~\cite{schuster2007ultrawideband} argue that \ac{AIC}~\cite{akaike1974new,ludden1994comparison,burnham2004multimodel,burnham2003model} is better suited for the purpose of choosing among fading distributions.
Later on, the \ac{AIC} was also used in~\cite{he2014vehicle,santos2010modeling,he2013measurements,guan2014propagation,he2013short}.
The \ac{AIC} can be seen as a form of Occam's razor as it penalizes the number of estimable parameters in the approximating model~\cite{burnham2003model} and hence aims for parsimony.

\subsection{Mathematical description of \ac{TWDP} fading}
An early form of \ac{TWDP} was analysed in~\cite{esposito1973statistical}. 
Durgin et al.~\cite{durgin2002new} introduced a random phase superposition formalism. 
Later,~\cite{rao2015mgf} achieved a major breakthrough and found a description of \ac{TWDP} fading as conditional Rician fading.
For the benefit of the reader, we will briefly repeat some important steps of~\cite{rao2015mgf}.

The \ac{TWDP} fading model in the complex-valued baseband is given as
\be 
r_\text{complex} = V_1 e^{j\phi_1}+V_2 e^{j\phi_2}+ X + jY~,  \label{eq:TWDPasSum}
\ee
where $V_1\ge 0$ and $V_2 \ge 0$ are the deterministic amplitudes of the non-fluctuating specular components. The phases $\phi_1$ and $\phi_2$ are independent and uniformly distributed in $(0,2\pi)$. The diffuse components are modelled via the law of large numbers as $X+jY$, where $X,Y\sim \mathcal{N}(0,\sigma^2)$.  
The $K$-factor is the power ratio of the specular components to the diffuse components 
\be 
K = \frac{V_1^2+V_2^2}{2\sigma^2}~. \label{eq:Kfact}
\ee
The parameter $\Delta$ describes the amplitude relationship among the specular components
\be
\Delta=\frac{2V_1 V_2}{V_1^2 + V_2^2}~.
\ee
The $\Delta$-parameter is bounded between $0$ and $1$ and equals $1$ iff both amplitudes are equal. 
The second moment of the envelope $r=|r_\text{complex}|$ of \ac{TWDP} fading is given as 
\be 
\mathbb{E}\big[r^2 \big]=\Omega=V_1^2+V_2^2+2\sigma^2~. \label{eq:Omega}
\ee
Expectation is denoted by $\mathbb{E}$.
For bounded amplitudes $V_1$ and $V_2$, a clever choice of $\sigma^2$ normalises $\Omega$, that is $\Omega \equiv 1$.
Starting from (\ref{eq:Omega}), by using (\ref{eq:Kfact}) we arrive at
\be
\sigma^2 = \frac{1}{2(1+K)}~.
\ee
Given the $K$ and $\Delta$ parameter ($\Omega\equiv 1$), the authors of \cite{kim2017comments} provide a formula for the amplitudes of both specular components
\begin{align}
V_{1,2} &= \frac{1}{2}\sqrt{\frac{K }{K+1}}\big( \sqrt{1+\Delta} \pm \sqrt{1-\Delta}	 \big) ~.
\end{align}

Real-world measurement data have $\Omega \neq 1$. 
To work with the formalism introduced above, we normalise the measurement data through estimating $\hat{\Omega}$ by the method of moments.
The second moment $\Omega$ of Rician fading and TWDP fading is merely a scale factor~\cite{lopez2016joint,lopez2016moment}. 
Notably, we are more concerned with a proper fit of $K$ and $\Delta$. 
Generally, estimation errors on $\Omega$ propagate to $K$ and $\Delta$ estimates.
However,~\cite{lopez2016joint} achieved an almost asymptotically efficient estimator with a moment-based estimation of $\Omega$.

Our envelope measurements are partitioned into 2 sets. 
We take the first set ($r_1, \ldots, r_n, \ldots r_N$) for parameter estimation of the tuple ($K,\Delta$) as described in Section~\ref{sec:par}, and hypothesis testing as described in Section~\ref{sec:validation}.
The first set is carefully chosen to obtain envelope samples that are approximately independent and identically distributed. 
The second set  ($r_1, \ldots, r_m, \ldots r_M$) is the complement of the first set.
We use the elements of the second set to estimate the second moment via
\be 
\hat{\Omega}=\frac{1}{M}\sum\limits_{m=1}^M r_m^2~, \label{eq:OmegaHat}
\ee
where $m$ is the sample index and $M$ is the size of the second set.
Partitioning is necessary to avoid biases through noise correlations of $\hat{\Omega}$ and ($\hat{K},\hat{\Delta}$)~\cite{goldberger1964econometric}.

\begin{figure}[H]
\centering
\includegraphics[width=0.65\textwidth]{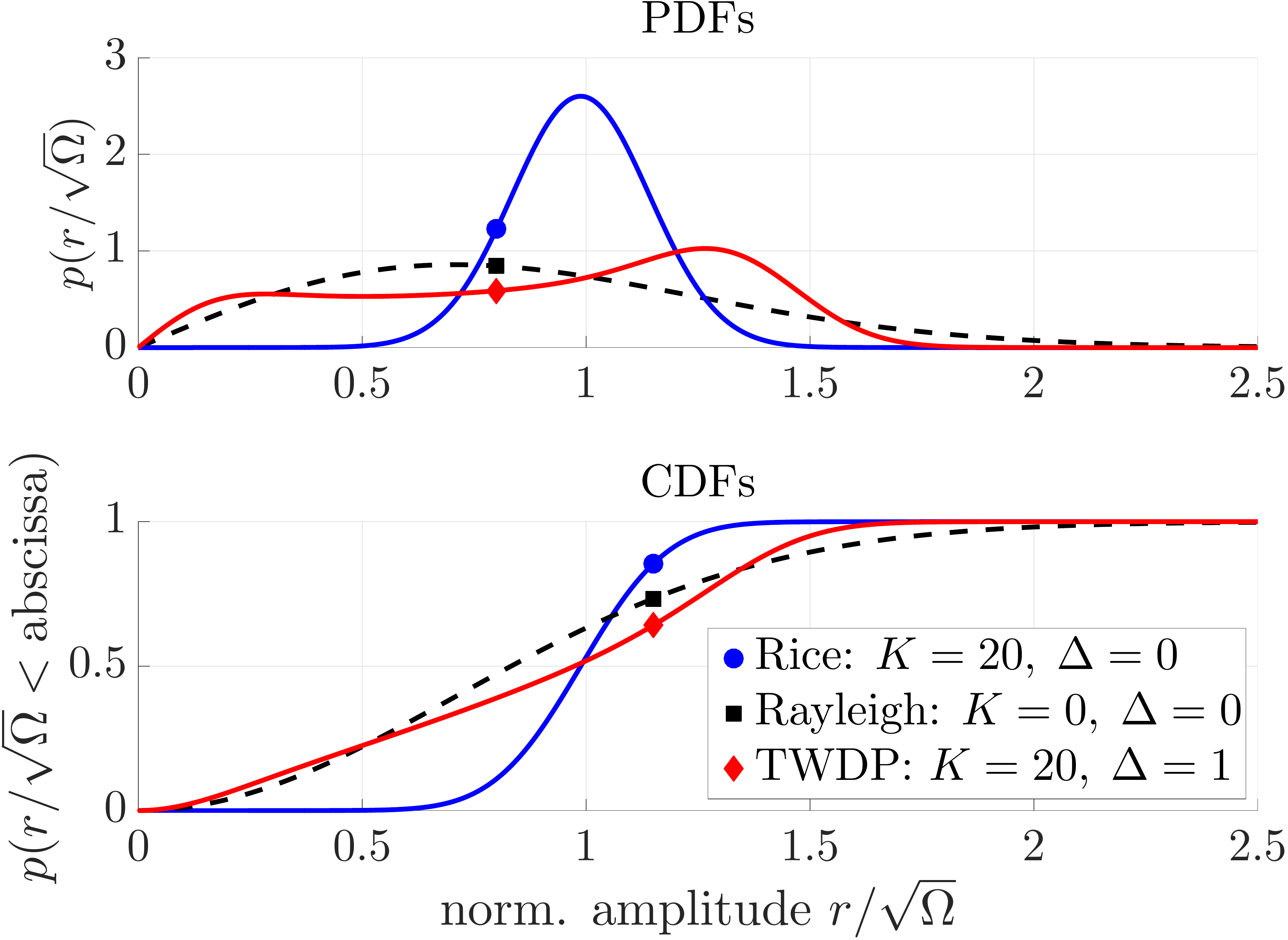}
\caption{{\bf Comparison of Rayleigh, Rician, and {TWDP} fading.}
The {TWDP} distribution with $\Delta=1$ deviates from the Rice distribution. {TWDP} fading's probability for deep fades is higher than for a Rayleigh distribution. \label{fig:TWDP_example}}
\end{figure}

By considering the estimate (\ref{eq:OmegaHat}) as true parameter $\Omega$, all distributions are parametrised by the tuple ($K,\Delta$), solely.
Example distributions are shown in Fig.~\ref{fig:TWDP_example}. 
The \ac{CDF} of the envelope of (\ref{eq:TWDPasSum}) is given in~\cite{rao2015mgf} as
\begin{align}
F_\text{TWDP}(r ; K, \Delta) =& \label{eq:CDF_TWDP} \\  1 - \frac{1}{2\pi}  \int\limits_{0}^{2\pi}  Q_1 \!\! &  \left( \sqrt{2 K \left[ 1+ \Delta \cos\left(\alpha\right)  \right]}, \frac{r}{\sigma}  \right) \notag \text{d}\alpha~. 
\end{align}
The Marcum Q-function is denoted by $Q_1(\cdot,\cdot)$.
For $\Delta \rightarrow 0$, Equation (\ref{eq:CDF_TWDP}) reduces to the well known Rice CDF
\be 
F_\text{Rice}(r ; K) = 1 -  Q_1 \left( \sqrt{2 K }, \frac{r}{\sigma}  \right)~.  \label{eq:CDF_Rice}
\ee

It might sound tempting to have a second strong radio signal present; in fact, however, two waves  can either superpose constructively or destructively and eventually lead to fading that is more severe than Rayleigh~\cite{frolik2007case,frolik2008appropriate,frolik2009compact,matolak2011worse,bakir2009diversity}. 
We observe the highest probability for deep fades for \ac{TWDP} fading in Fig.~\ref{fig:TWDP_example}.

\subsection{Parameter estimation and model selection}
\label{sec:par}

Note that our model of \ac{TWDP} fading (\ref{eq:TWDPasSum}) does (obviously) not contain noise. 
Over our wide frequency range (in \ac{MC1} we have $7$\,GHz bandwidth) the receive noise power spectral density is not equal. 
A statistical noise description that is valid over our wide frequency range is frequency-dependent.
To avoid the burden of frequency-dependent noise modelling, we only take measurement samples which lie at least $10$\,dB above the noise power and ignore noise in our estimation.

\begin{figure}[H]
\centering
\includegraphics[width=0.65\textwidth]{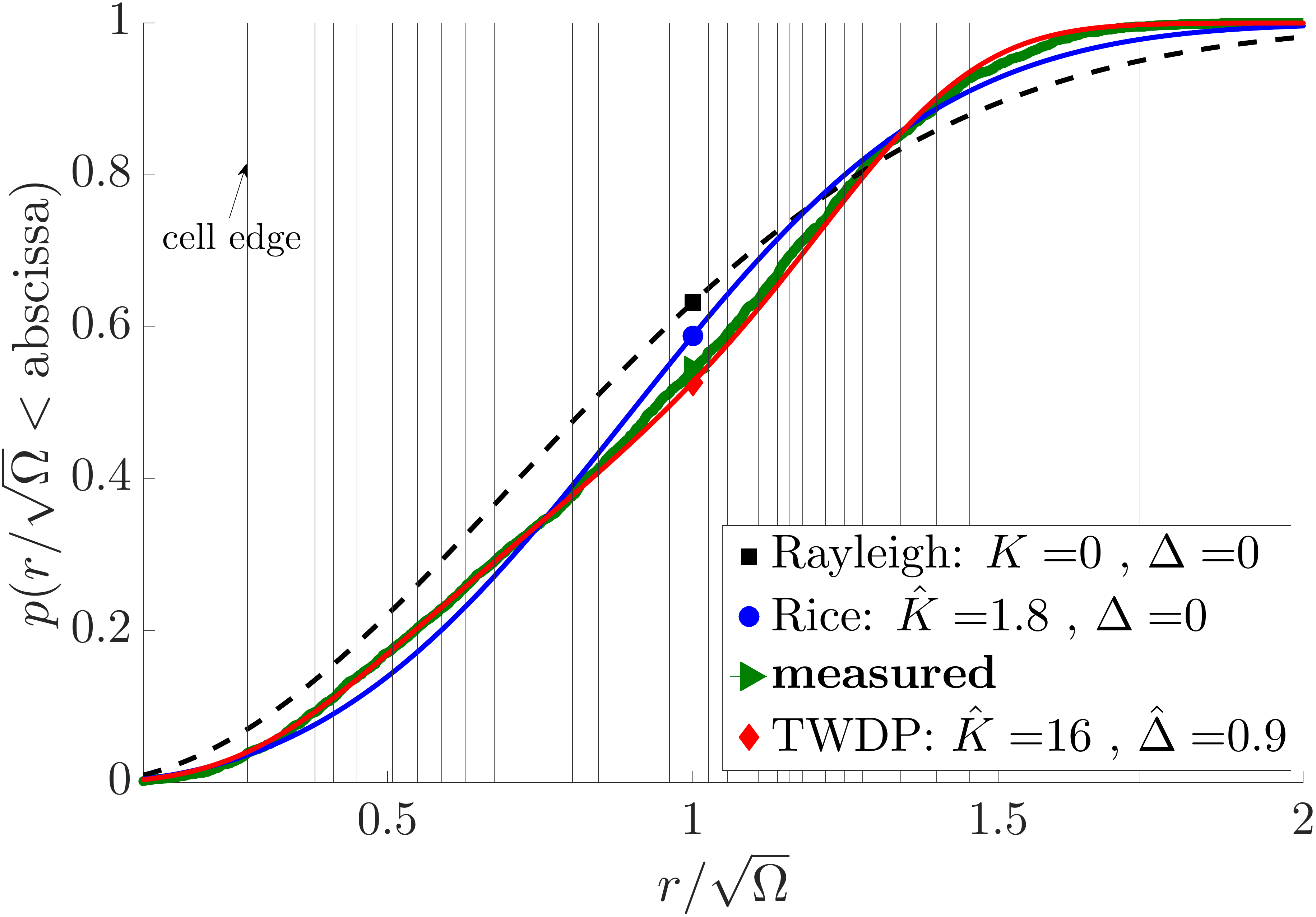}
\caption{{\bf CDF: Distribution fitting for exemplary frequency domain measurement data.}
Illustration of the maximum likelihood fitted Rice distribution and the  maximum likelihood fitted {TWDP} fading distribution. The Rician $K$-factor and the {TWDP} $K$-factor deviate significantly. Rayleigh fading is plotted as reference. }\label{fig:cdf}
\end{figure}

Having the envelope measurement data set ($r_1, \ldots, r_n, \ldots r_N$) at hand, we are seeking a distribution of which the observed  realisations $r_n$ appear most likely.
To do so, we estimate the parameter tuple ($K$,$\Delta$) via the maximum likelihood procedure
\begin{align}
(\hat{K},\hat{\Delta}) &=\argmax_{K,\Delta} \sum_{n=1}^N \operatorname{ln} \frac{\partial F_\text{TWDP}(r_n ; K, \Delta)}{\partial r} \notag\\
						 &=\argmax_{K,\Delta} \sum_{n=1}^N\operatorname{ln} f_\text{TWDP}(r_n; K, \Delta)  \notag\\ 							
						&=\argmax_{K,\Delta} \sum_{n=1}^N \operatorname{ln} \mathcal{L}(K, \Delta | r_n)~.  \label{eq:loglik}          
\end{align}
We denote the \ac{PDF} by $f(\cdot)$, $n$ denotes the sample index, and $N$ the size of the set.
To solve (\ref{eq:loglik}), we first discretise $K$ and $\Delta$ in steps of $0.05$.  Next, we calculate $\frac{\partial F_\text{TWDP}(r ; K, \Delta)}{\partial r}$ for all parameters via numerical differentiation.
Within this family of distributions, we search for the parameter vector maximizing the log-likelihood function~(\ref{eq:loglik}). 
For the optimal Rice fit, the maximum is searched within the parameter slice ($K, \Delta = 0$).
An exemplary fit of Rician and \ac{TWDP} fading is shown in Fig.~\ref{fig:cdf}.
As a reference, Rayleigh fading ($K=0, \; \Delta=0$) is shown as well.

To select between Rician fading and \ac{TWDP} fading, we employ Akaike's information criterion~(AIC). 
The \ac{AIC} is a rigorous way to estimate the Kullback–Leibler divergence, that is, the relative entropy based on the maximum-likelihood estimate~\cite{burnham2003model}.
Given the maximum-likelihood fitted parameter tuple ($\hat{K}, \hat{\Delta}$) of \ac{TWDP} fading and Rician fading, we calculate the sample size corrected \ac{AIC}~\cite[p. 66]{burnham2003model} for Rician fading ($\text{AIC}_\text{R}$) or TWDP fading ($\text{AIC}_\text{T}$)
\begin{align}
&\hspace{-5ex}\text{AIC}_\text{R/T} =  \label{eq:AIC} \\&\hspace{-5ex}- 2 \operatorname{ln} \mathcal{L}_\text{R/T}(\hat{K}, \hat{\Delta}|r)  + 2U_\text{R/T} + \frac{2U_\text{R/T} (U_\text{R/T}+1)}{N-U_\text{R/T}-1}~, \notag
\end{align}
where $U$ is the model order.
For Rician fading the model order is $U_\text{R}=1$, since we estimate the $K$-factor, only.
For \ac{TWDP} fading  $U_\text{T}=2$, as $\Delta$ is estimated additionally.
The second moment $\Omega$ (estimated already with a different data set before the parameter estimation) is not part of the ML estimation (\ref{eq:loglik}) and therefore not accounted in the model order $U$.
We choose between Rician fading and \ac{TWDP} fading based on the lower \ac{AIC}.

\subsection{Validation of the chosen model}
\label{sec:validation}

Based on (\ref{eq:AIC}), one of the two distributions, Rice or TWDP, will always yield a better fit.
To validate whether the chosen distributions really explains the data, we state the following statistical hypothesis testing problem:
\begin{align}
\mathcal{H}_0: &\begin{cases}
 F_\text{Rice}(r; \hat{K}), & \text{if }\text{AIC}_\text{R} \le \text{AIC}_\text{T}  \\
 F_\text{TWDP}(r; \hat{K}, \hat{\Delta}), &\text{else}
\end{cases} \notag \\
\mathcal{H}_1:   &\begin{cases}
\neg F_\text{Rice}(r; \hat{K}),  &\text{if }\text{AIC}_\text{R} \le \text{AIC}_\text{T}  \\
\neg F_\text{TWDP}(r; \hat{K},\hat{\Delta}), &\text{else}
\end{cases} \label{eq:nullhypo}
\end{align}
The Boolean negation is denoted by $\neg$.
Our statistical tool is the g-test~\cite{mcdonald2009handbook,woolf1957log}\footnote{The well known chi-squared test approximates the g-test via a local linearisation~\cite{hoey2012two}.}.
At a significance level $\alpha$, a null hypothesis is rejected if
\begin{equation}
G=2\sum_{i=1}^{m}  O_i  \ln \bigg( \frac{O_i}{E_i} \bigg) \overset{  \mathbf{?}}{>} \chi_{(1-\alpha, m-e)}^2~,
\end{equation}
where $O_i$ is the observed bin count in cell $i$ and $E_i$ is the expected bin count in cell $i$ under the null hypothesis $\mathcal{H}_0$. 
The cell edges are illustrated with vertical lines in Fig.~\ref{fig:cdf}. 
The cell edges are chosen, such that $10$ observed bin counts fall into one cell. 
The estimated parameters of the model are denoted by $e$.
For Rician fading we estimate $e=2$ ($\Omega, K$) parameters and for \ac{TWDP} fading we estimate $e=3$ ($\Omega, K, \Delta$) parameters in total.
The $(1-\alpha)$\,--\,quantile of the chi-square distribution with $m - e$ degrees of freedom is denoted by $\chi_{(1-\alpha, m-e)}^2$. 
The prescribed confidence level is $1-\alpha=0.01$~.

\section{Floor plan and set-ups for \ac{MC1} and \ac{MC2}}
\label{sec:mech}

\begin{figure}[H]
\centering
\includegraphics[width=0.65\textwidth]{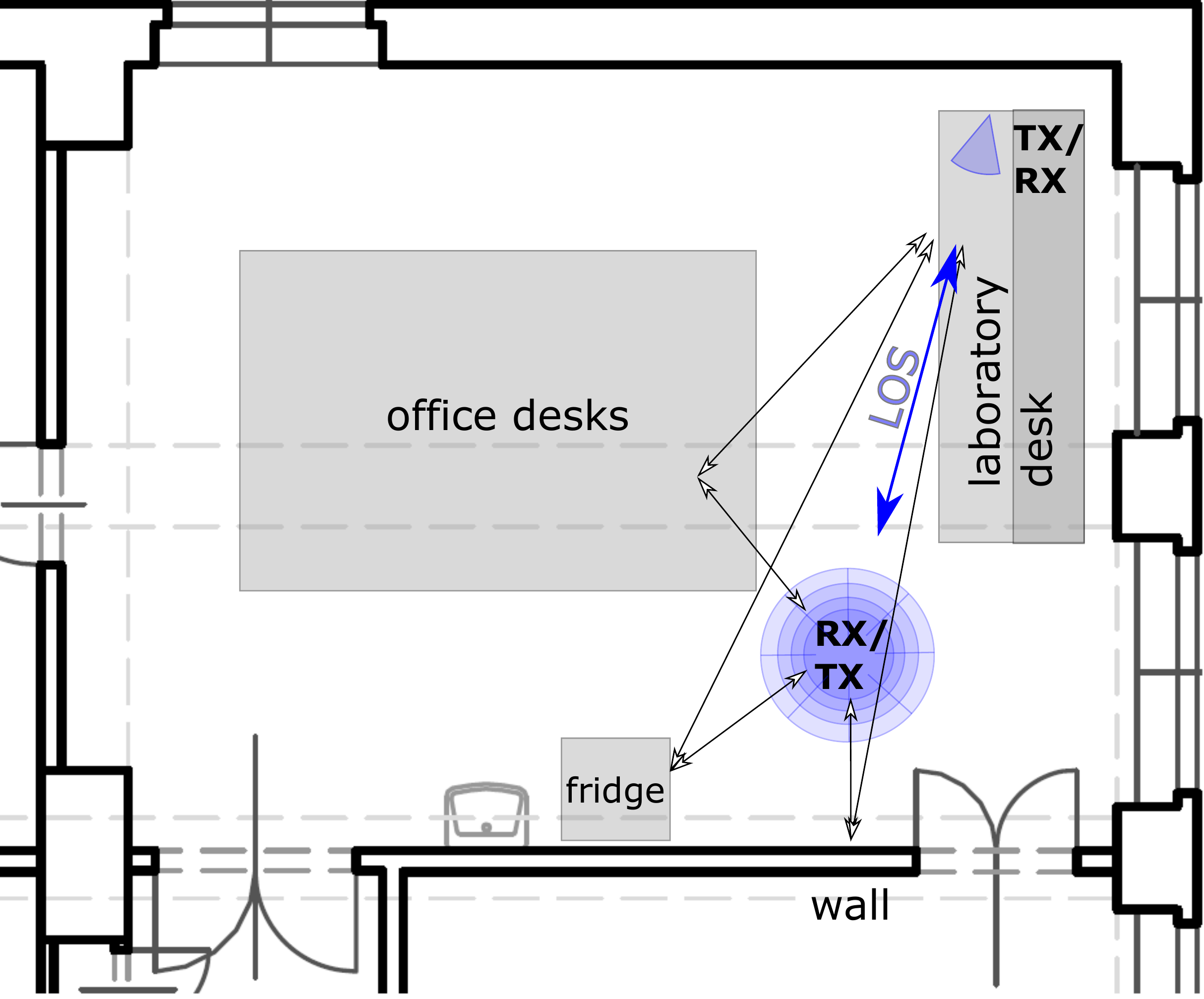}
\caption{{ \bf Floor plan of the measured environment.}
      The floorplan indicates the multipath components that are visible in the measurement results.
      {TX} and {RX} switch roles for {MC2}. 
      {TX}/{RX} in the right upper corner of the room is always static. 
      {RX}/{TX} in the middle of the room is steerable, indicated by the spider's web.  \label{fig:floor} }
\end{figure}

Our measured environment is a mixed office and laboratory room. 
There are office desks in the middle of the room and at the window side, there are laboratory desks, see Fig~\ref{fig:floor}.
The main interacting objects in our channel are office desks, a metallic fridge, a wall, and the surface of the laboratory desk.
These objects are all marked in Fig.~\ref{fig:floor}.

\begin{figure}[H]
\centering
\includegraphics[width=0.65\textwidth]{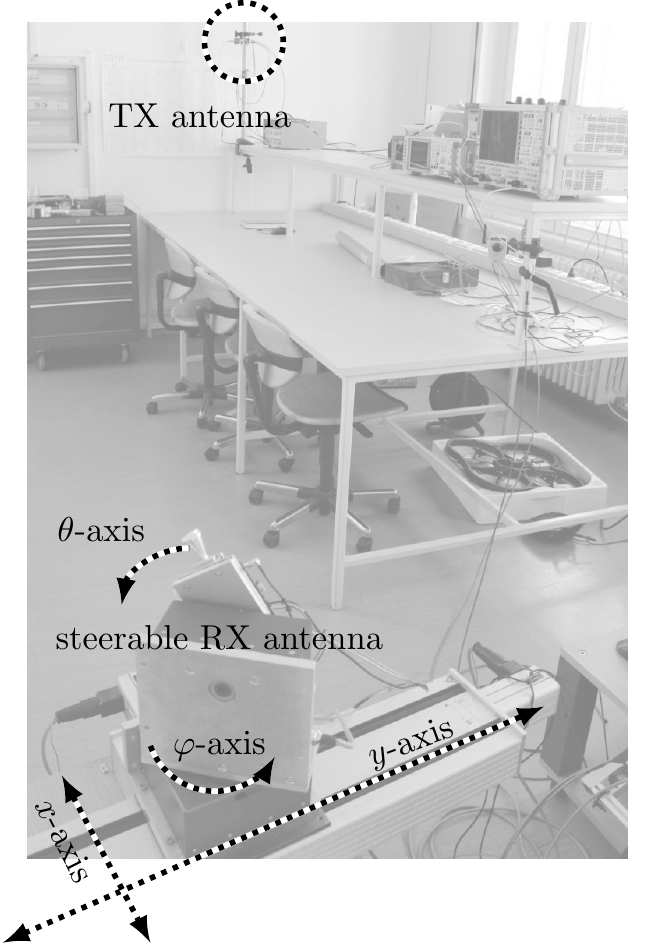}
\caption{{\bf Photograph of the mechanical set-up for {MC1} from the receiver point of view.}
      The receive antenna, a conical $20$\,dBi horn, is mounted on a multi-axis positioning and rotating system. The azimuthal and elevation angle are controlled to scan the whole upper hemisphere. The multi-axis system moves and rotates the horn antenna such that its phase center stays at the same $(x,y)$ coordinate during the directional scan.  \label{fig:mech_1}}
\end{figure}

Our directional measurements are carried out by using the traditional approach of mechanically steered directional antennas~\cite{spencer2000modeling,durgin2003wideband}.
As directional antennas, $20$\,dBi conical horn antennas with an $18^\circ$  $3$\,dB opening angle are used.
Our polarisation is determined by the LOS polarisation. When TX and RX are facing each other at LOS, the polarisation is co-polarised and the E-field is orthogonal to the floor.
In \ac{MC1}, the essential mechanical adaptation to the state-of-the-art directional channel sounding set-up~\cite{Fuschini2017,vehmas2016millimeter} is that the elevation-over-azimuth positioner is mounted on an xy-positioning stage. 
Thereby, we compensate for all linear translations caused by rotations and keep the phase center of the horn antenna always at the same $(x,y)$ coordinate, see Fig.~\ref{fig:mech_1}.
The $z$ coordinate is roughly $70$\,cm above ground but varies $13$\,cm for different elevation angles.

For \ac{MC2} we add another linear guide along the $z$-axis to compensate for all introduced offsets. 
The horn antenna's phase center is thereby lifted upwards by one metre.
Now we are able to fix the phase center of the horn antenna at a specific $(x,y,z)$ coordinate in space.
The whole mechanical set-up and the fixed phase center is illustrated in Fig.~\ref{fig:mech_2}.

\section{\ac{MC1}: Scalar-valued wideband measurements}
\label{sec:scalar}

A wireless channel is said to be small-scale fading, if the \ac{RX} cannot distinguish between different \acp{MPC}. 
Depending on the position of the \ac{TX}, the position of the \ac{RX} and the position of the interacting components, the \acp{MPC} interfere constructively or destructively~\cite[pp. 27]{molisch2012wireless}. 
The fading concept only asks for a single carrier frequency, whose \acp{MPC} arrive with different phases at the \ac{RX}. 
By spatial sampling a statistical description of the fading process is found.

In \ac{MC1}, the spatial $(x,y)$\,--\,coordinate (of \ac{TX} and \ac{RX}) is kept constant. 
Different phases of the impinging \acp{MPC} are realised by changing the \ac{TX} frequency over a bandwidth of $7$\,GHz. 
Thereby we implicitly rely on frequency translations to estimate the moments of the spatial fading process.

\subsection{Measurement set-up}
\begin{figure}[H]
\centering
\includegraphics[width=0.65\textwidth]{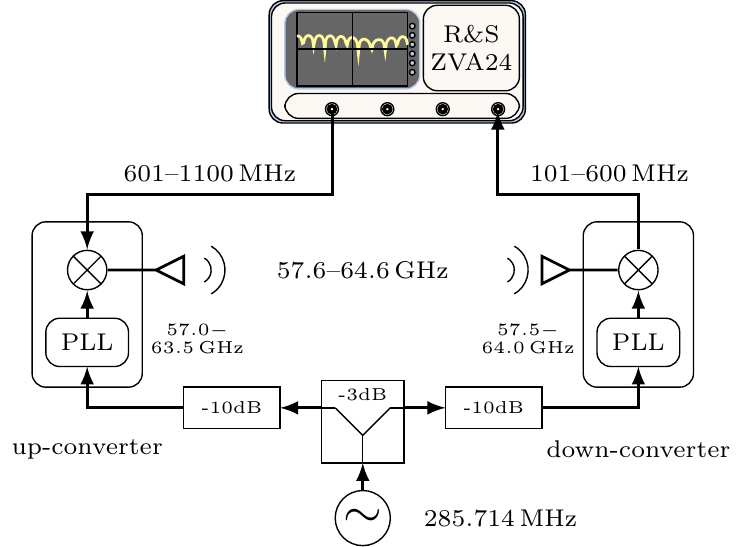}
\caption{{ \bf {RF} set-up for {MC1}.}
The combination of different PLL scaling factors allows for a measurement bandwidth of $7$\,GHz.
The reference clock for the up-converter and the down-converter is shared.
The power splitter has an isolation of $30$\,dB.
To avoid possible leakage on the clock distribution network, attenuators additionally decouple both converters. 
The transfer function is measured applying the conversion gain (mixer) measurement option of the R\&S VNA. \label{fig:RFscalar}
}
\end{figure}

We measure the forward transfer function with an Rohde and Schwarz R\&S ZVA24 \ac{VNA}. 
The \ac{VNA} can measure directly up to $24$\,GHz. 
For \ac{mmWave} up-conversion and down-conversion, we employ modules from Pasternack~\cite{pasternack}. 
They are based on radio frequency integrated circuits described in~\cite{zetterberg2015open}.  
The up-converter module and the down-converter module are operating built-in synthesizer \acp{PLL}, where the \ac{LO} frequency is calculated as
\be 
f_\text{LO}=\nicefrac{7}{4} \cdot s_\text{PLL} \cdot 285.714\,\text{MHz}\approx s_\text{PLL} \cdot 500\,\text{MHz}  ~.
\ee
The scaling factor of the synthesizer \ac{PLL} counters is denoted by $s_\text{PLL}$.
For $f_\text{LO} \approx 60\,\text{GHz}$, the scaling factor is $s_\text{PLL} = 120$.
To avoid crosstalk, we measure the transfer function via the conversion gain (mixer) measurement option of our \ac{VNA} and operate the transmitter and receiver at different baseband frequencies: $601$ to $1100$\,MHz and $101$ to $600$\,MHz. 
The set-up is shown in Fig.~\ref{fig:RFscalar}.

\subsection{Receive power and fading distributions}

\begin{figure}[H]
\centering
\includegraphics[width=0.65\textwidth]{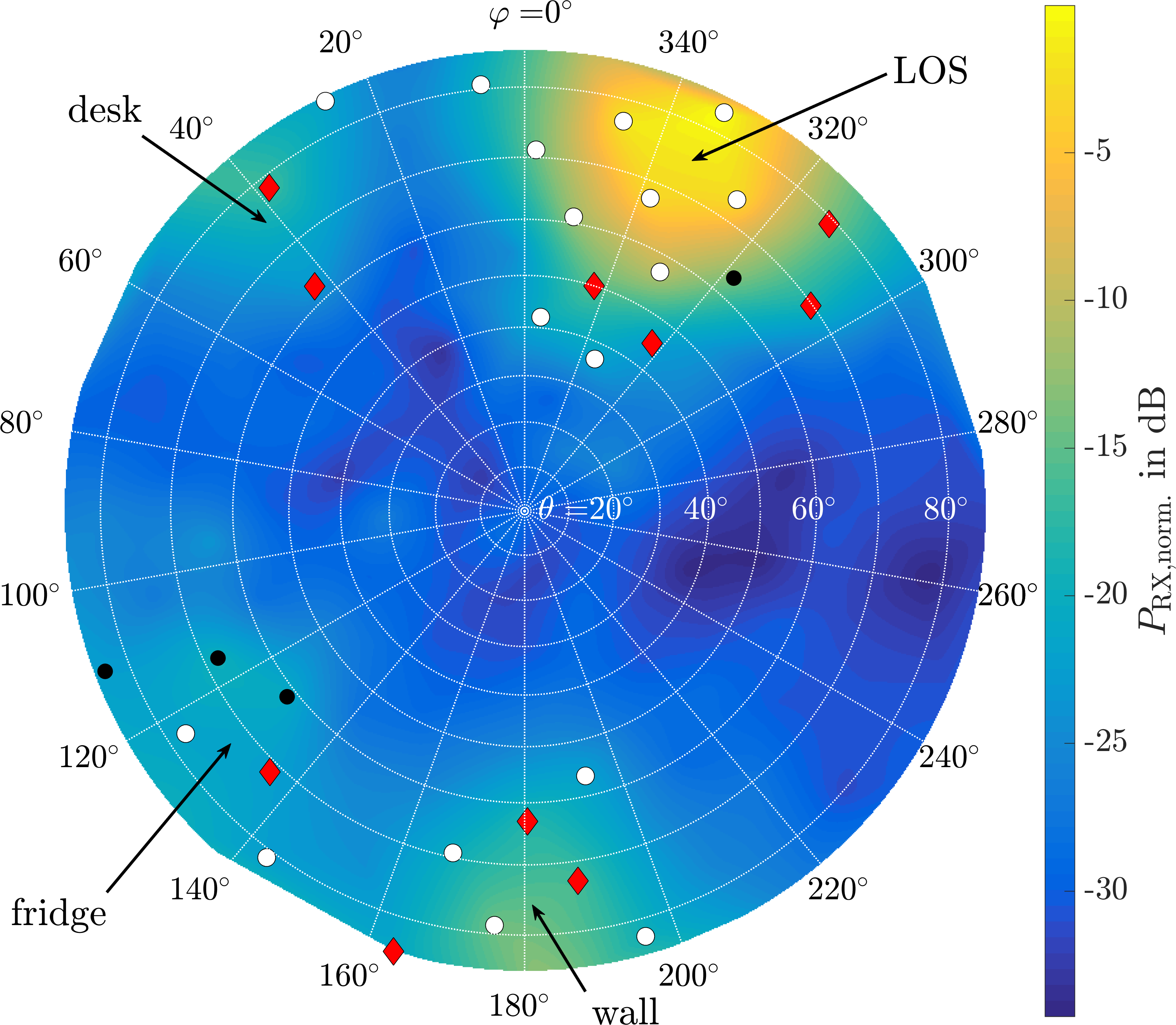}
\caption{{\bf Estimated directional receive power of {MC1}. \label{fig:scaler_pow}}
     There are four main interacting objects leading to stronger receive power (marked in the figure). TWDP fading occurs whenever the \ac{LOS}-link is not perfectly aligned or the reflecting structure is not perfectly plain. Red diamonds mark {TWDP} fading and white circles mark Rician fading. Black markers show points where both distributions are rejected by the hypothesis test. Directions less than $10$\,dB above the noise level are not evaluated. }
\end{figure}

In Fig.~\ref{fig:scaler_pow}, we show  the estimated received mean power of $7$\,GHz bandwidth, normalised to the maximum \ac{RX} power, that is
\be
P_\text{RX,norm.}(\varphi,\theta)=\frac{\hat{\Omega}(\varphi,\theta)}{\max_{\varphi',\theta'}(\hat{\Omega}(\varphi',\theta'))}~.
\ee
As already mentioned in Section~\ref{sec:meth}, we partition the frequency measurements into two sets.
The normalised receive power is calculated according to (\ref{eq:OmegaHat}), with frequency samples spaced by $2.5$\,MHz.
Every tenth sample is left out as these samples are used for fitting of ($K,\Delta$) and hypothesis testing.
We display the results via a stereographic projection from the south pole and use $\tan (\nicefrac{\theta}{2})$ as azimuthal projection. 
All samplings points, lying at least $10$\,dB above the noise level, are subject of our study.
They are displayed with red, white or black markers.
Sampling points where we decided for \ac{TWDP} fading, following the procedure described in Section~\ref{sec:meth}, are marked with red diamonds.
White circles mark points for which \ac{AIC} favours Rician fading.
Four points are marked black. 
These points failed the null hypothesis test and we neither argue for Rician fading nor  for \ac{TWDP} fading.
\ac{TWDP} fading occurs whenever the \ac{LOS}-link is not perfectly aligned or if the interacting object cannot be described by a pure reflection.

\begin{figure}[H]
\centering
\includegraphics[width=0.65\textwidth]{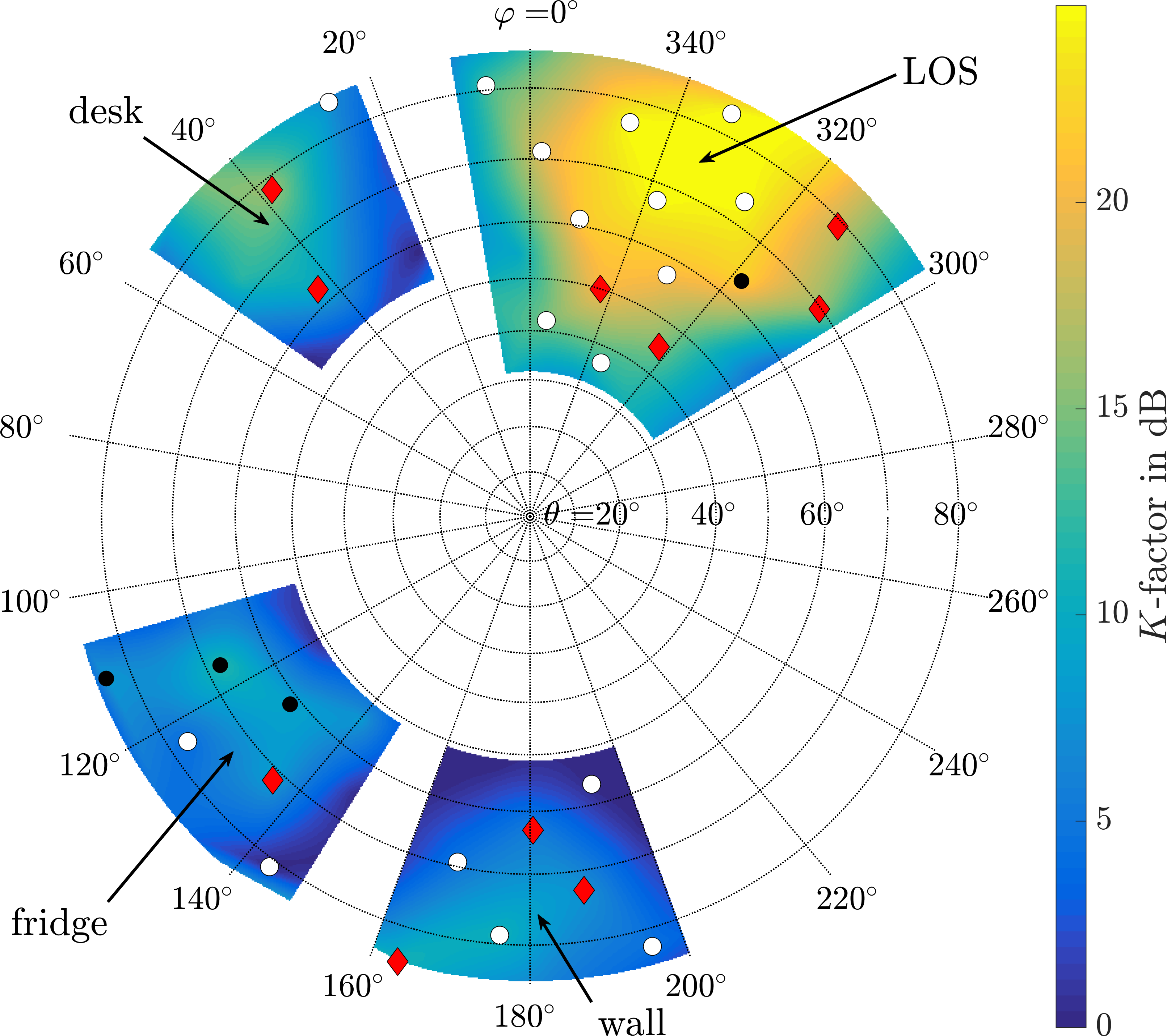}

\vspace{2ex}

\includegraphics[width=0.65\textwidth]{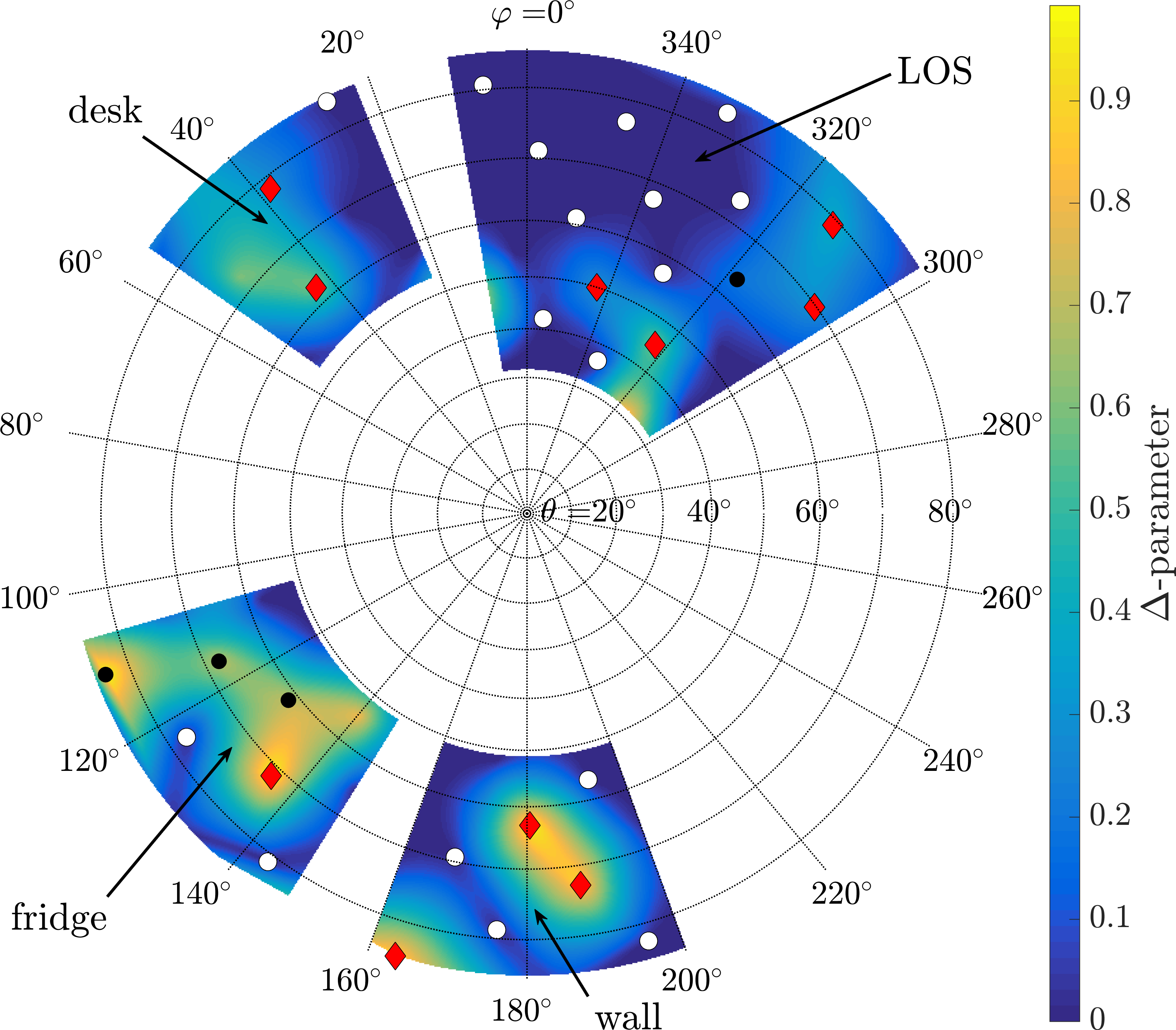}
  \caption{{\bf Estimated $K$-factor and $\Delta$-parameter  of {MC1}.}
      We plot the $K$-factor estimate of the selected hypothesis. The $K$-factor behaves analogously to the {RX} power.  At LOS, the $K$-factor is far above $20$\,dB.
The desk reflection has a surprisingly high $K$-factor of about $15$\,dB.
Other reflections have $K$-factors of approximately $10$\,dB. The $\Delta$-parameter for reflections tends to be close to $1$. Markers have the same meaning as in Fig.~\ref{fig:scaler_pow}. \label{fig:K_and_D_scalar}}
      \end{figure}

\begin{figure}[H]
\centering
\includegraphics[width=0.65\textwidth]{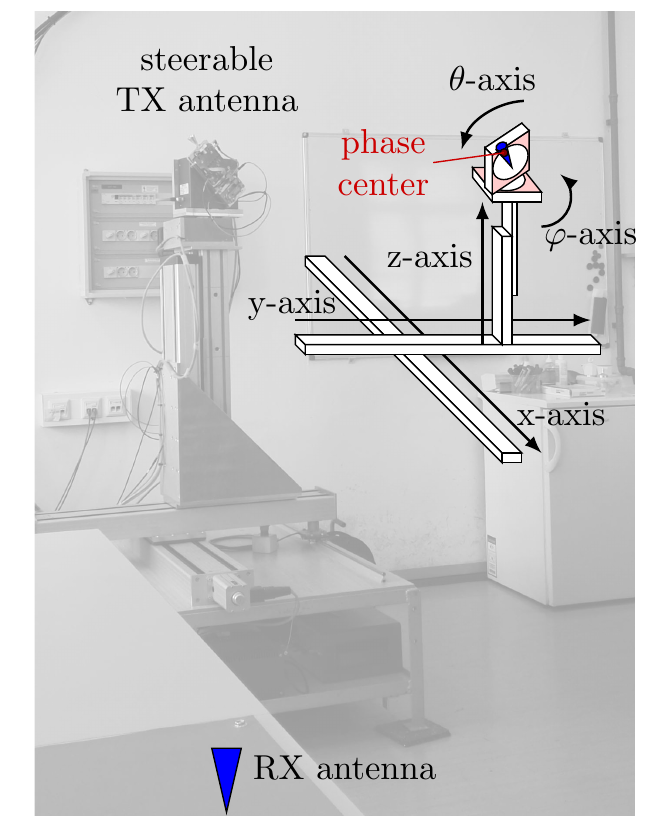}
\caption{{\bf Photograph of the improved mechanical set-up for {MC2} from the receiver point of view.}
      Our mechanical set-up consists now of five independent axis to fully compensate all offsets introduced by rotation. A schematic sketch is superimposed. All five axis are necessary to rotate the horn antenna  around the phase center at a fixed  $(x,y,z)$ coordinate. Notice that TX and RX switch roles as compared to Fig.~\ref{fig:mech_1}.  \label{fig:mech_2}}
\end{figure}

In Fig.~\ref{fig:K_and_D_scalar}, the $K$-parameter of the selected hypothesis is illustrated. 
Figure~\ref{fig:K_and_D_scalar} shows either the Rician K-factor or the \ac{TWDP} K-factor, depending on the selected hypothesis.
Note that their definitions are fully equivalent.
For Rician fading, the amplitude $V_2$ in (\ref{eq:TWDPasSum}) is zero by definition.
Whenever the \ac{RX} power is high, the $K$-factor is high.
Below the $K$-estimate, the estimate of $\Delta$ is shown.
Here again, by definition, $\Delta=0$ whenever we decide for Rician fading.
For interacting objects, the parameter $\Delta$ tends to be close to one. 
Note, that decisions based on \ac{AIC} select \ac{TWDP} fading mostly when $\Delta$ is above $0.3$.
Smaller $\Delta$ values do not change the distribution function sufficiently to justify a higher model order.

\section{MC2: Vector-valued spatial measurements}
\label{sec:vec}

In contrast to \ac{MC1}, we no longer rely on frequency translations and are indeed sampling the channel in space.
The fading results we present in Section~\ref{sec:fading_complex} are evaluated at a single frequency. 
Fading is hence determined by the obtained spatial samples, exclusively.

\subsection{Measurement set-up}

\begin{figure}[H]
\centering
\includegraphics[width=0.65\textwidth]{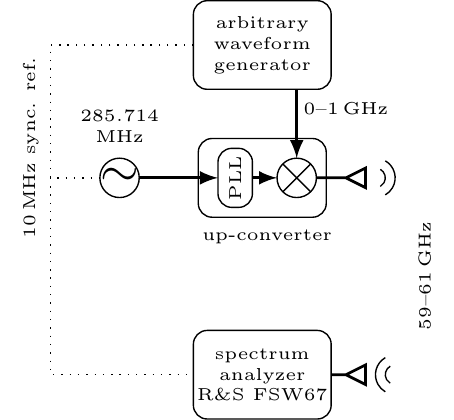}
  \caption{{\bf RF set-up for MC2.}   
      The VNA from Fig.~\ref{fig:RFscalar} is replaced with an AWG and an SA. This leads to a set-up where we obtain phase information as well. An option of the SA gives us direct access to the baseband IQ samples.  \label{fig:RFvector}}
\end{figure}

At the transmitter side,  a $2$\,GHz wide waveform is produced by an \ac{AWG}. 
A multi-tone waveform (OFDM) with Newman phases~\cite{sangodoyin2012real,kim2013development,TUW-268504} is applied as sounding signal. 
The signal has $401$ tones (sub-carriers) with a spacing of $5$\,MHz. 
This large spacing assures that our system is not limited by phase noise~\cite{TUW-259766}. 
The \ac{TX} sequence is repeated $2\,000$ times to obtain a coherent processing gain of $33$\,dB for i.i.d. noise. 
The Pasternack up-converter (the same as in \ac{MC1}) shifts the baseband sequence to $60$\,GHz. 
The $20$\,dBi conical horn antenna, together with the up-converter is mounted on a five axis positioner to directionally steer them.
As receiver, a \ac{SA} (R\&S FSW67) with a 2 GHz analysis bandwidth is used. 
The received \ac{IQ} baseband samples are obtained from the \ac{SA}. 
The whole system is sketched in Fig.~\ref{fig:RFvector}.
 
In \ac{MC2}, for feasibility reasons, \ac{TX} and \ac{RX} switch places compared to \ac{MC1}.
The \ac{RX} in form of the \ac{SA} is put onto the laboratory table.
The \ac{RX} $20$\,dBi conical horn antenna is directly mounted at the \ac{RF} input of the \ac{SA}.
The \ac{SA} is located on a table close to a corner of the room; the \ac{RX} antenna is not steered.

Similar to the set-ups of \cite{nissel2017lowlatency,lerch2014vienna,caban2011synchronization,laner2011time}, proper triggering between the arbitrary waveform generator and the \ac{SA} ensures a stable phase between subsequent measurements.
 
\subsection{Receive power} 

\begin{figure}[H]
\centering
\includegraphics[width=0.65\textwidth]{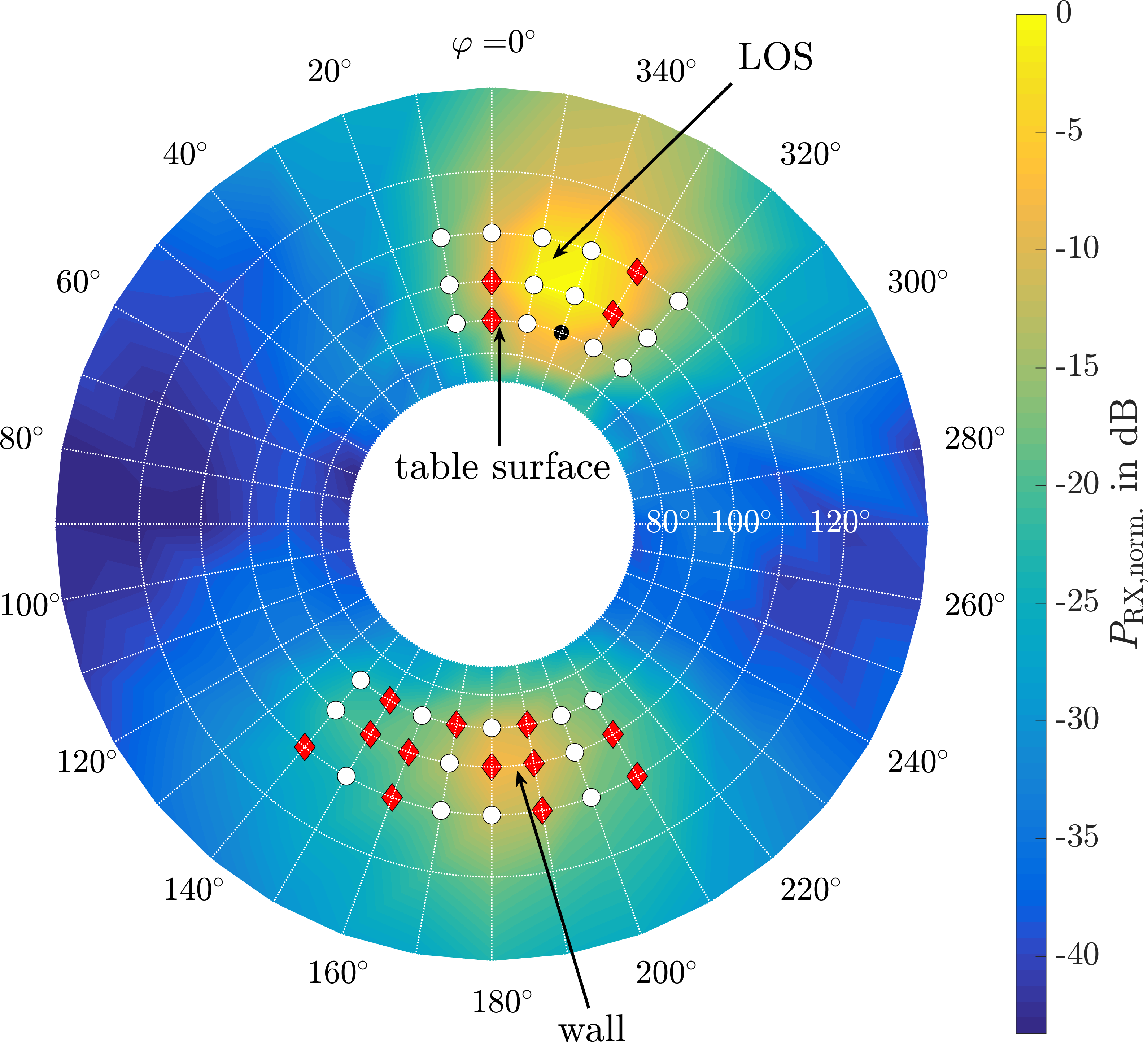}
  \caption{{\bf Estimated directional receive power of {MC2}.}
      Due to the elevated position of the steerable horn antenna, two interacting objects from Fig.~\ref{fig:scaler_pow}, namely the desk and the fridge, are no longer visible. LOS and the wall reflection are still present. These regions are the only ones which are spatially sampled. Markers have the same meaning as in Fig.~\ref{fig:scaler_pow}.\label{fig:P_vector}}
\end{figure}

For the calculation of the \ac{RX} power, averaged over $2$\,GHz bandwidth, we perform a sweep through azimuth and elevation at a single coordinate.
The \ac{LOS} and wall reflection from \ac{MC1} are still visible in Fig.~\ref{fig:P_vector}.
Fading is evaluated at a single frequency in the subsection below.
Nevertheless, we already indicate fading distributions by markers in Fig.~\ref{fig:P_vector} in order to better orient ourselves later on.

As the steerable horn antenna is above the office desks and the fridge level, these interacting objects do not become apparent. 
In case the steerable \ac{TX} does not hit the \ac{RX} at \ac{LOS} accurately, the table surface acts as reflector and a \ac{TWDP} model explains the data.
For wall reflections, with non-ideal alignment, \ac{TWDP} also explains the data best.

\subsection{Fading distributions} 
\label{sec:fading_complex}

\begin{figure}[H]
\centering
\includegraphics[width=0.65\textwidth]{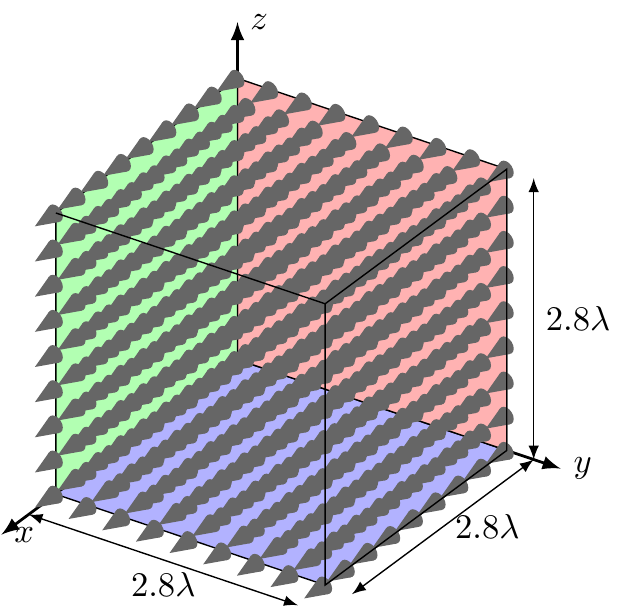}
\caption{{\bf Spatial sampling grid.} For one specific direction, we draw $9\times 9 \times 9=729$ samples uniformly from a cube of side length $2.8\lambda$. The distance between samples is $0.35\lambda$ with a repeat accuracy of $\pm 0.004\lambda$. The orientation of the horn antenna is indicated via the cone shape at the sampling points.}
\label{fig:sampling}
\end{figure}

To obtain different spatial realisations, with the horn antenna pointing into the same direction, the coordinate of the apparent phase center is moved to ($x,y,z$)\,--\,positions uniformly distributed within a cube of side length $2.8\lambda$, see Fig.~\ref{fig:sampling}.
We realise a set of $9 \times 9 \times 9 = 729$ directional measurements.
This results in a spacing between spatial samples of $0.35\lambda$ in each direction. 
Although $\nicefrac{\lambda}{2}$ sampling is quite common~\cite{dupleich2017investigations,samimi201628}, we choose the sampling frequency to be co-prime with the wavelength, to circumvent periodic effects~\cite{haining1993spatial}.
We restrict our spatial extend to avoid changes in large-scale fading.
Only at directions with strong reception levels do we perform spatial sampling\footnote{Spatial sampling for all directions takes more than three days.}.
Similar as in the previous section, we partition the measurements into 2 sets. 
The partitioning is made according to a 3D chequerboard pattern. 
The first set is used for the estimation of the second moment $\hat{\Omega}$ and the second set is used for the parameter tuple ($K,\Delta$).

\begin{figure}[H]
\centering

\includegraphics[width=0.65\textwidth]{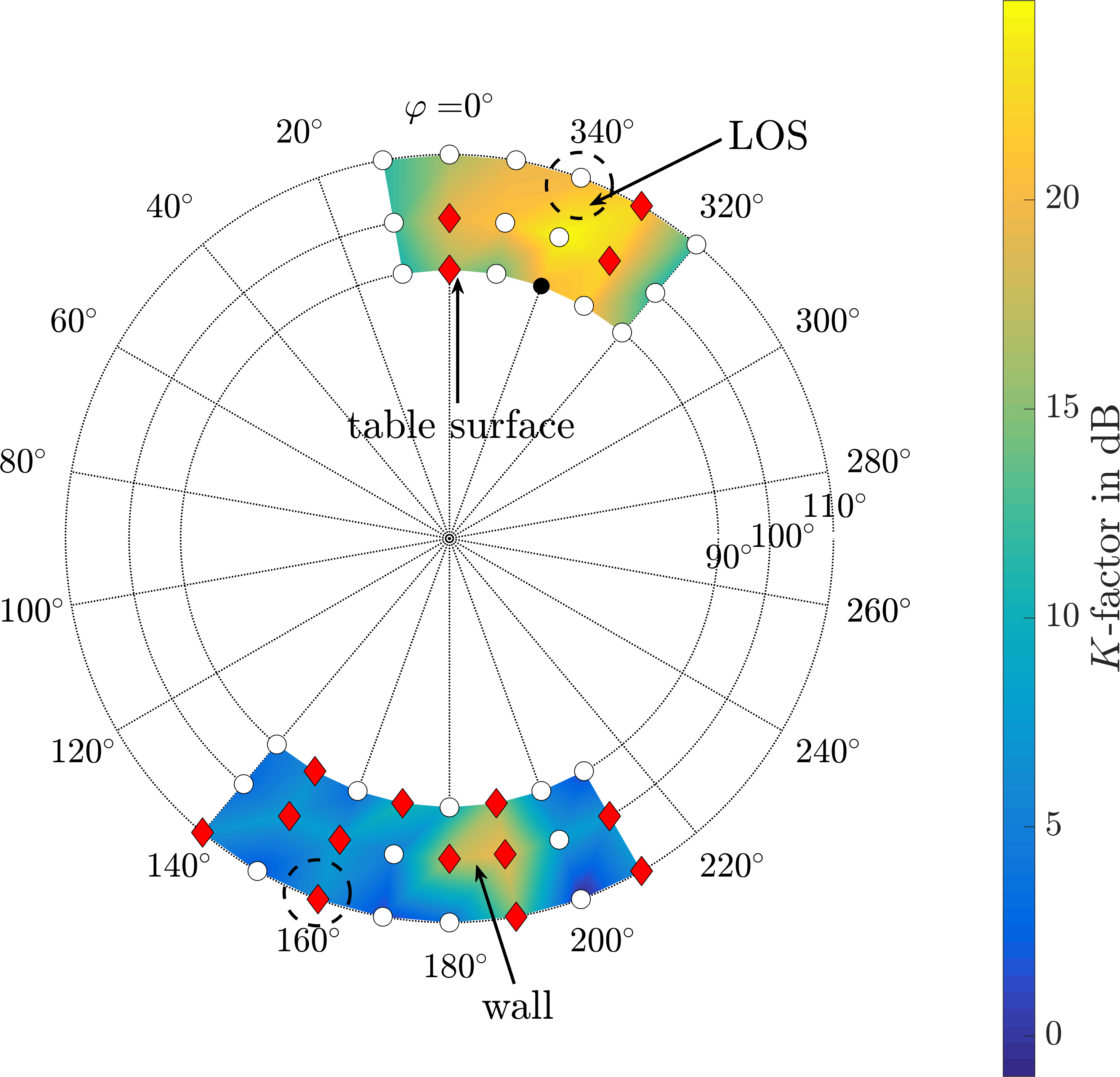}

\vspace{2ex}

\includegraphics[width=0.65\textwidth]{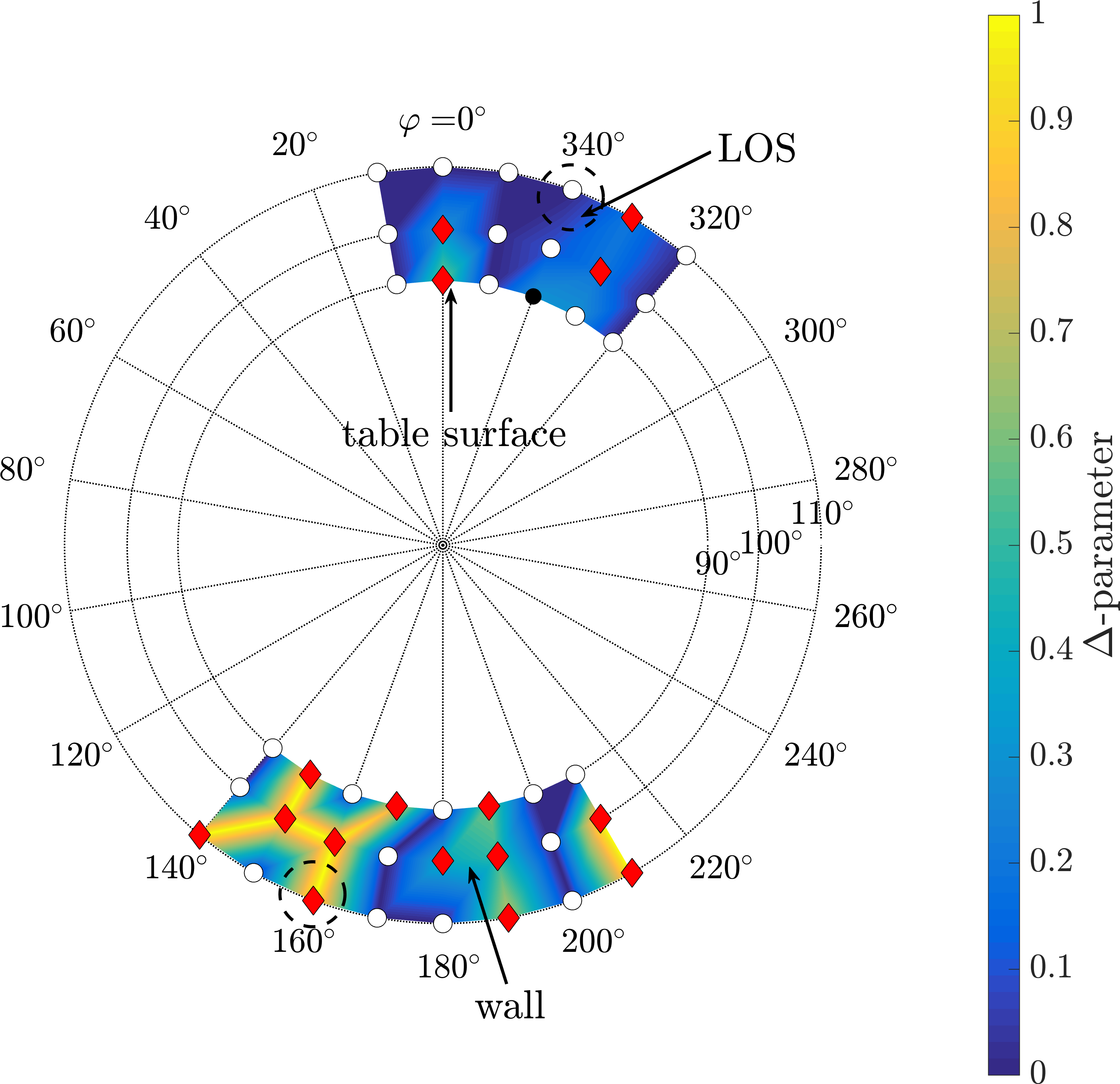}
  \caption{{\bf Estimated $K$-factor and $\Delta$-parameter of {MC2}.}
      Due to the elevated position, the wall reflection has a $6$\,dB increased $K$-factor as compared to MC1, see top part of Fig.~\ref{fig:K_and_D_scalar}. If the beam is not perfectly aligned, Rician fading turns again into TWDP fading.   Wall reflections described by \ac{TWDP} fading have a $\Delta$-parameter of close to one. The table surface reflection leads to a significantly smaller reflected component ($\Delta \ll 1$). The encircled sampling points  are subject of further study in Sections~\ref{sec:spatial} and \ref{sec:CIR}. Markers have the same meaning as in Fig.~\ref{fig:scaler_pow}. \label{fig:K_and_D_vector}}
\end{figure}

The best fitting $K$-factors, in both regions with strong reception, are illustrated in Fig.~\ref{fig:K_and_D_vector}, top part. 
Below the $\Delta$-parameters are provided.
Remember, the \ac{RX} in form of an \ac{SA} is put on the laboratory table.
In case the \ac{TX} is not perfectly aligned, a reflection from the table surface yields a fading statistic captured by the \ac{TWDP} model.
The interaction with the wall,  similar to Fig.~\ref{fig:K_and_D_scalar}, has again regions best modelled via \ac{TWDP} fading.

\section{\ac{MC2}: Efficient computation of the spatial correlation}
\label{sec:spatial}

The wall reflection from the previous section is now subject to a more detailed study.
Our spatial samples are used to show spatial correlations among the drawn samples.

Our three-dimensional sampling problem, see again Fig.~\ref{fig:sampling}, is treated via two-dimensional slicing.
For the calculation of the spatial (2D) autocorrelation function, we apply the Wiener--Khintchine--Einstein theorem,  that relates the autocorrelation function of a wide-sense-stationary random process to its power spectrum~\cite{khintchine1934korrelationstheorie}. In two dimensions, this theorem reads~\cite{ell2000hypercomplex,moxey2003hypercomplex}
\begin{equation}
\mathcal{F}_{2D}\left\lbrace \Mat{C}(x,y)\right\rbrace =  S(x',y')~,
\end{equation}
where $\Mat{C}$ is the 2D-autocorrelation and $S$ is the power spectral density of a 2D signal. 
The operator $\mathcal{F}_{2D}$ denotes the 2D Fourier transform.
We calculate all 2D autocorrelation functions $\Mat{C}^{(z,f)}$ of one $x-y$ slice at height $z$ at a single frequency $f$ through
\begin{align}
&\mathcal{F}_{2D}\big\lbrace  \Mat{C}^{(z,f)}(x,y) \big\rbrace =  \mathcal{F}_{2D} \big\lbrace \Re\big\lbrace\Mat{H}^{(z,f)}\big(x',y'\big) \big\rbrace \big\rbrace  \notag \\ & \hspace{7ex} \odot\operatorname{conj} \big\lbrace \mathcal{F}_{2D} \big\lbrace \Re\big\lbrace\Mat{H}^{(z,f)}\big(x',y'\big)\big\rbrace \big\rbrace \big\rbrace~. \label{eq:auto_orig}
\end{align}
The symbols $\odot$ denotes the Hadamard multiplication. 
The operator $\operatorname{conj} \lbrace \cdot \rbrace$ denotes complex conjugation.
To ensure a real-valued autocorrelation matrix (instead of a generally complex representation~\cite{moxey2003hypercomplex}), from the complex-valued channel samples  only the real parts $\Re\lbrace \cdot \rbrace$ are taken. 
The spatial autocorrelation of the imaginary parts are identical. 
One could also analyse the magnitude and phase individually. 
While the correlation of the magnitude stays almost at 1, the phase correlation patterns are similar to those of the real part. 

The 2D Fourier transform $\mathcal{F}_{2D}$ is realised via a 2D \ac{DFT}.
The 2D \ac{DFT} is calculated via a multiplication with the \ac{DFT} matrix $\Mat{D}$ from the left and the right.
To mimic a linear convolution with the \ac{DFT}, zero padding is necessary.
We hence take the matrix ${\widetilde{\Mat{H}}}^{(z,f)}$
\begin{equation}
\widetilde{\Mat{H}}^{(z,f)} = \left( \begin{array}{cc}
\Re\lbrace\Mat{H}^{(z,f)} \rbrace& \Mat{0} \\ 
\Mat{0} & \Mat{0}
\end{array} \right)~. \label{eq:H_construct}
\end{equation}
Furthermore, the finite spatial extend of our samples acts as rectangular window.
The rectangular window leads to a triangular envelope of the the autocorrelation function.
This windowed spatial correlation is denoted by
\begin{align}
\Mat{C}_{\text{windowed}}^{(z,f)} =  \Mat{D}^H   \bigg( & \big(\Mat{D}  {\widetilde{\Mat{H}}}^{(z,f)} \Mat{D} \big) \label{eq:auto_window} \\ & \odot \operatorname{conj} \big\lbrace \Mat{D} {\widetilde{\Mat{H}}}^{(z,f)} \Mat{D} \big\rbrace  \bigg)  \Mat{D}^H   \notag.
\end{align}
To compensate the windowing effect, we calculate the spatial correlation of the rectangular window, constructed in accordance to (\ref{eq:H_construct})
\begin{align}
\Mat{S} =  \Mat{D}^H   \bigg( & \bigg(\Mat{D}  \bigg( { \begin{array}{cc}
\mathds{1} & \Mat{0} \\ 
\Mat{0} & \Mat{0}\notag
\end{array} } \bigg) \Mat{D} \bigg) \\ & \odot \operatorname{conj} \bigg\lbrace \Mat{D} \bigg( { \begin{array}{cc}
\mathds{1} & \Mat{0} \\ 
\Mat{0} & \Mat{0}\notag
\end{array} } \bigg) \Mat{D} \bigg\rbrace  \bigg)  \Mat{D}^H~.
\end{align}
The matrix $\mathds{1}$ denotes the all-ones matrix.
Matrix $\Mat{S}$ compensates the truncation effect of the autocorrelation through element-wise (Hadamard) division, denoted by $\oslash$.
Finally, the efficient computation of the spatial correlation (\ref{eq:auto_orig}) reads
\begin{align}
\Mat{C}^{(z,f)} =  \Mat{C}_{\text{windowed}}^{(z,f)} \oslash \Mat{S}~. \label{eq:auto}
\end{align}

\begin{figure}[H]
\centering

{\bf Wall:}

\includegraphics[width=0.65\textwidth]{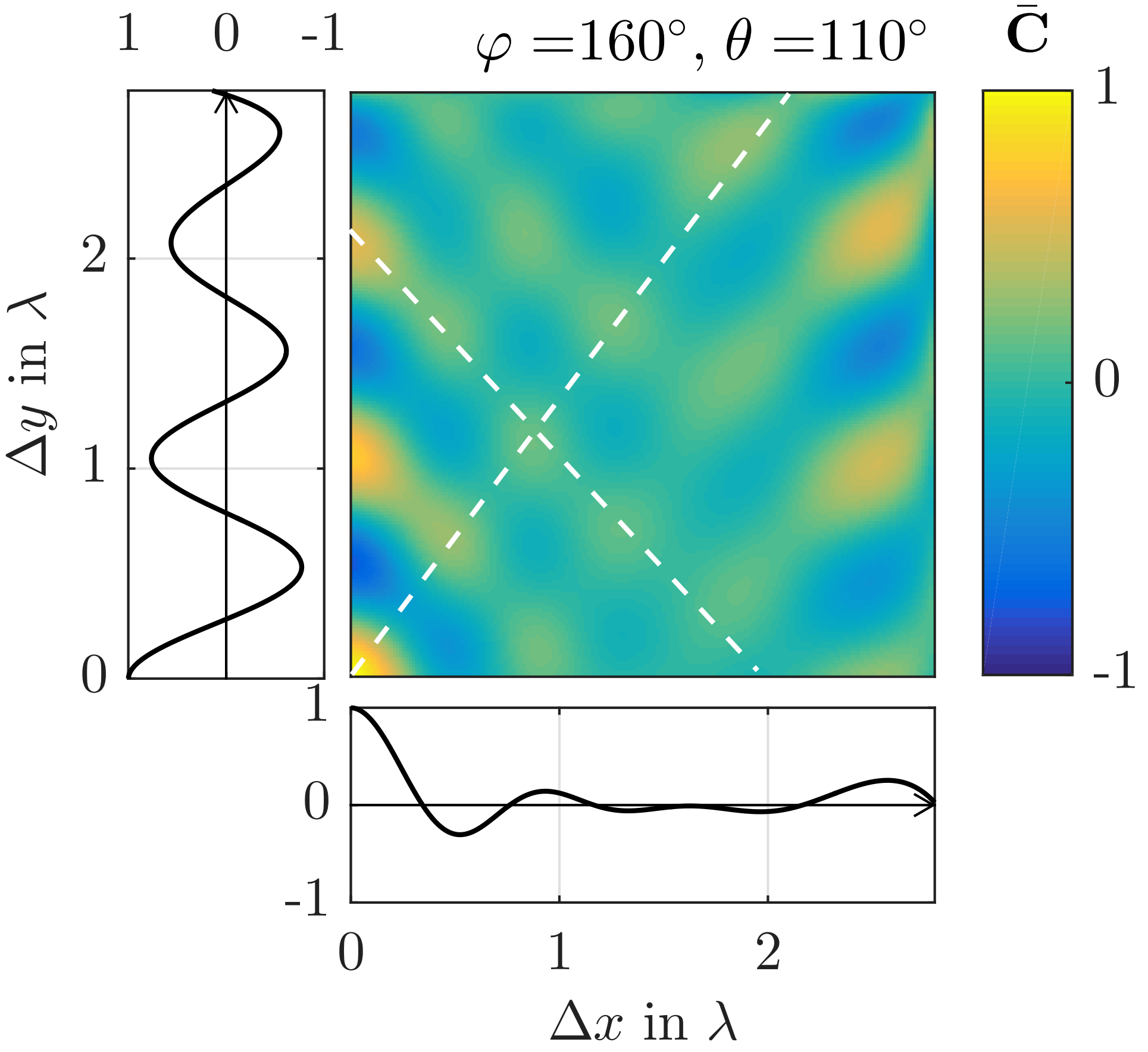}

\vspace{2ex}

{\bf LOS:}

\includegraphics[width=0.65\textwidth]{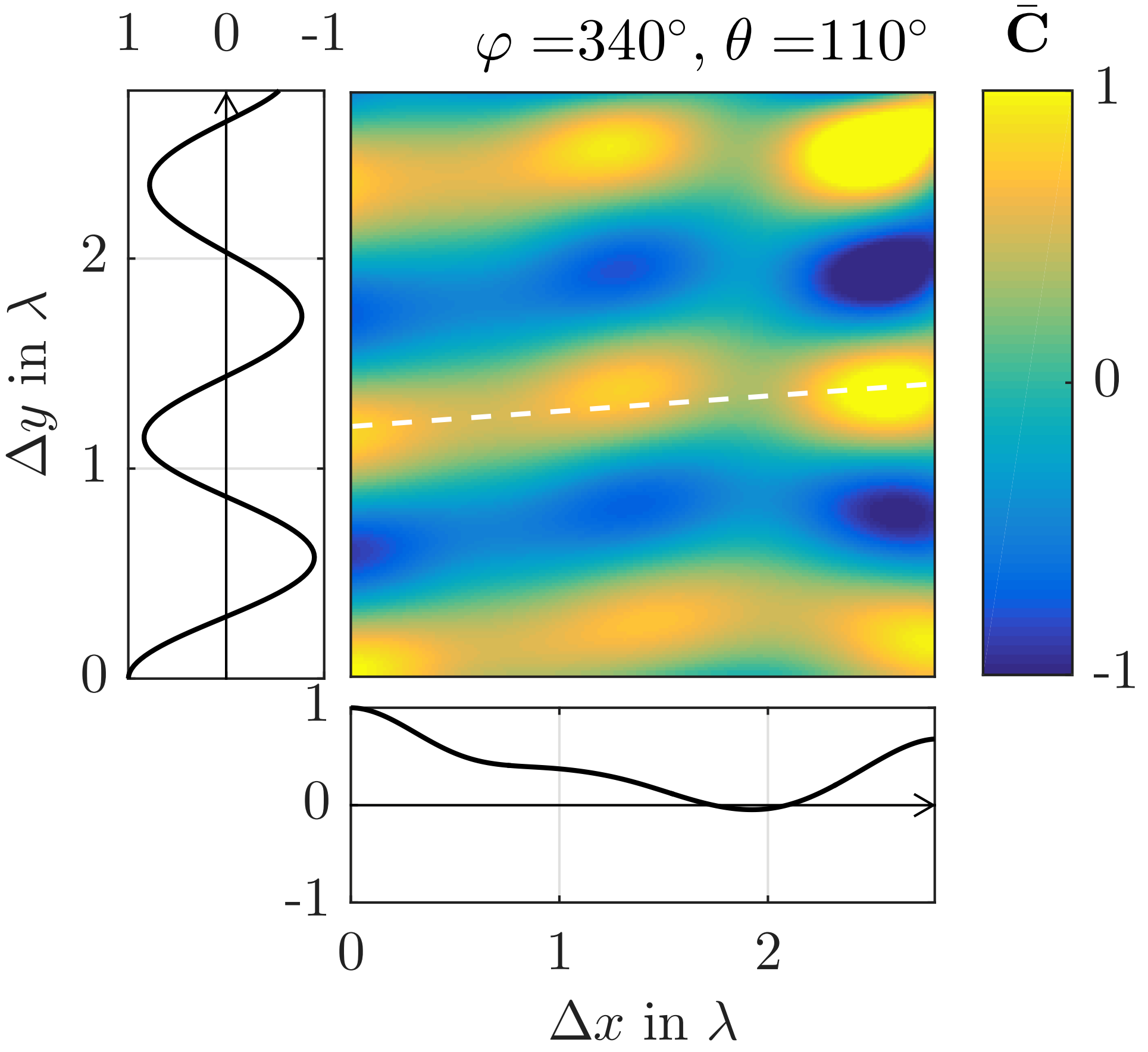}

\caption{{\bf Spatial correlation plot at $\varphi=160^\circ, 340^\circ$ and $\theta=110^\circ$.}
For the wall reflection at $\varphi=160^\circ$, the pattern shows an interference  of two plane waves, supporting the TWDP fading assumption.  For LOS at $\varphi=340^\circ$, we observe a spatial correlation pattern dominated by one wave.  The white dashed lines illustrate   plane wave phase fronts.   \label{fig:spat1}}
\end{figure}

At a distance of $0.35\lambda$, the measurement data is still correlated, therefore we are able to view our correlation results on the finer, interpolated grid.
The interpolation factor is $20$.
That means that we calculate our spatial correlations on a grid of $\nicefrac{0.35\lambda}{20}=0.0175\lambda$ distance.
The very efficient implementation of~(\ref{eq:auto}) is applied to all (parallel) 2D slices and to all frequencies. 
All realisations in $z$ and $f$ are averaged
\be 
\bar{\Mat{C}}= \frac{1}{9} \frac{1}{401}  \sum_{z=1}^9 \sum_{f=1}^{401}  \Mat{C}^{(z,f)}~.
\ee
Furthermore, we plot one-dimensional autocorrelation functions, evaluated along $x$ and $y$, together with their two-dimensional representations.
We provide two spatial correlation plots evaluated at an azimuth angle of $\varphi=340^\circ$ and $\varphi=160^\circ$ in Fig.~\ref{fig:spat1}, both at an elevation angle of $\theta=110^\circ$. 
The top part of Fig.~\ref{fig:spat1} shows a correlation pattern dominated by a single wave.
The spatial correlation below shows an interference pattern, which is intuitively explained by a superposition of two plane waves.
The one dimensional correlations, evaluated either at the $x$-axis or at the $y$-axis, show this oscillatory  behaviour as well.

\section{MC2: Time gated fading results}
\label{sec:CIR}

\begin{figure}[H]

\centering

{\bf Wall:}

\vspace{1ex}

\includegraphics[width=0.65\textwidth]{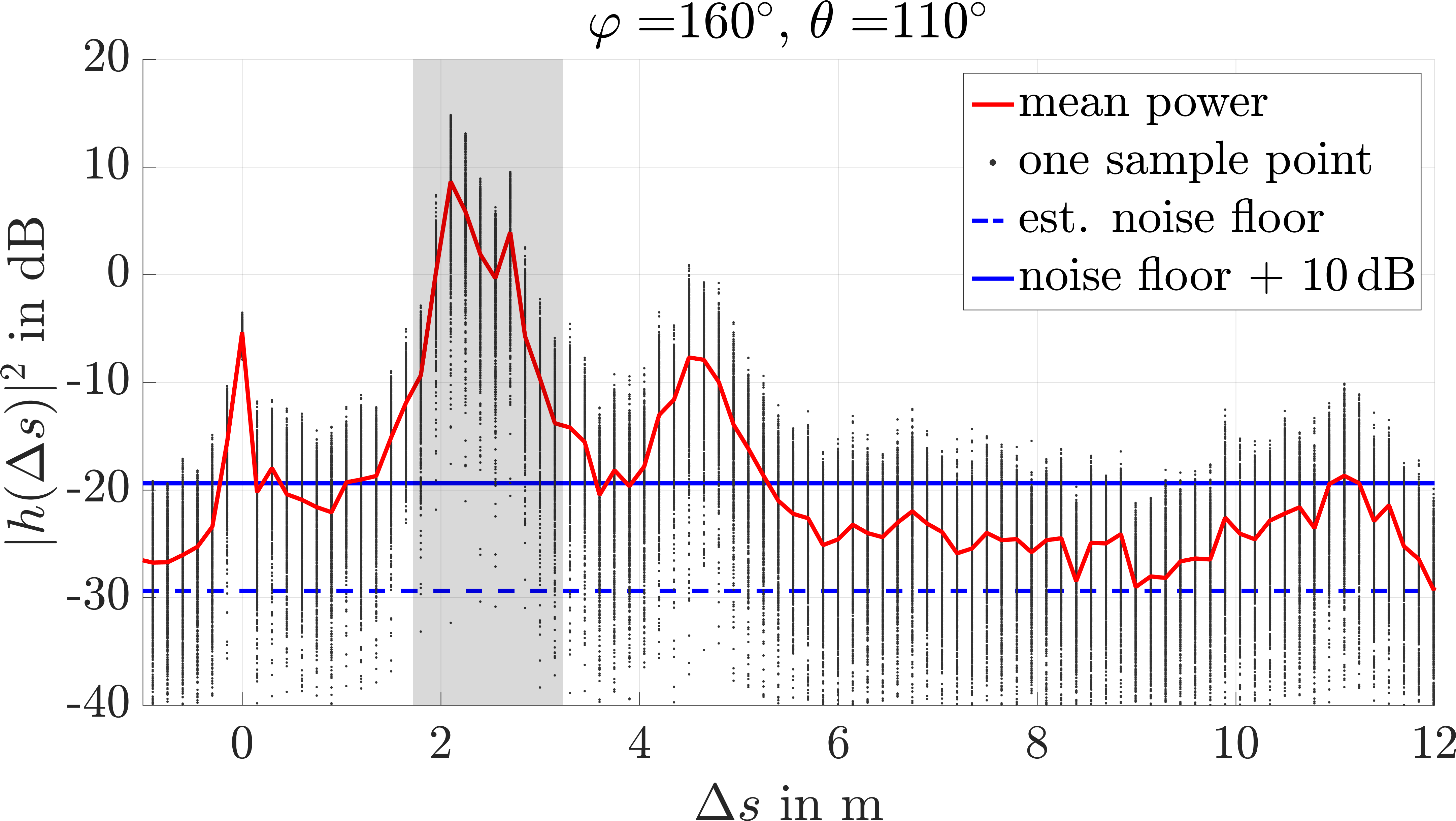}

\vspace{2ex}

{\bf LOS:}

\vspace{1ex}

\includegraphics[width=0.65\textwidth]{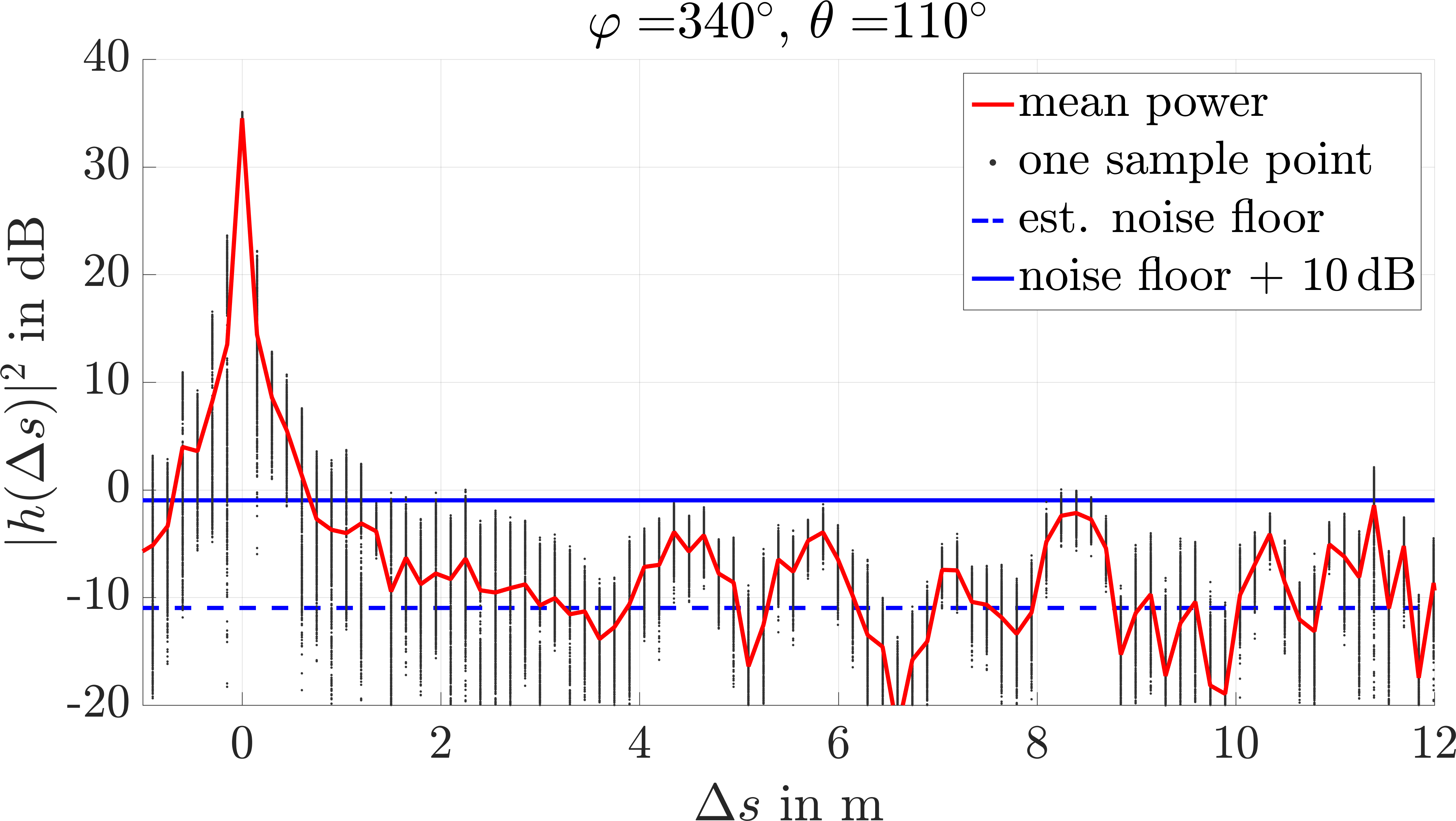}
\caption{{\bf Scatter-plot of the CIRs.}
	We plot the CIRs as a function of spatial distance, where $\Delta s=0$ corresponds to the LOS distance. Our spatial resolution (a channel tap) is $15$\,cm. 
	The spatial extend of our sampled cube (729 samples) is $2.8\lambda=1.4$\,cm, a magnitude smaller than the spatial resolution.
         The scatter-plot is evaluated at a wall reflection ($\varphi=160^\circ$) and at {LOS} ($\varphi=340^\circ$).
      The mean power  is plotted with a continous red line. We observe that the arrival cluster centred at $2.5$\,m fades very deeply. 
      The gray highlighted region around $2.5$\,m is further analysed in Fig.~\ref{fig:CIRzoom}. \label{fig:CIR}}
\end{figure}

To confirm that our observations are not artefacts of our measurement set-up, for example back-lobes of the horn antenna, we now study the wireless channel in the time domain. 
Our $2$\,GHz wide measurements from \ac{MC2} allow for a time resolution of approximately $0.5$\,ns.
This corresponds to a spatial resolution of $15$\,cm.
We plot the \acp{CIR} as a function of distance, namely the \ac{LOS} excess length $\Delta s$, that is 
\be
h(\Delta s) = h\big( \big( \tau - \tau_\text{LOS} \big)c_0 \big)~.
\ee
The scatter-plot of the \acp{CIR} for $\varphi=160^\circ$ is shown in Fig.~\ref{fig:CIR}. 
The \ac{LOS} \ac{CIR} at $\varphi=340^\circ$ is displayed as reference as well.
The steerable \ac{TX} is positioned more than a metre apart from the wall. 
This amounts in an excess distance of approximately two to three metres. 
At this excess distance, a cluster of multipath components is present. 
Note, if the horn antenna points towards the wall, the wave emitted by the back-lobe of the horn antenna is received at zero excess distance.
Still, the receive power of the back-lobe is far below the components arriving from the wall reflection.
Fading is hence determined by the wall scattering behaviour.

\begin{figure}[H]
\centering
\includegraphics[width=0.65\textwidth]{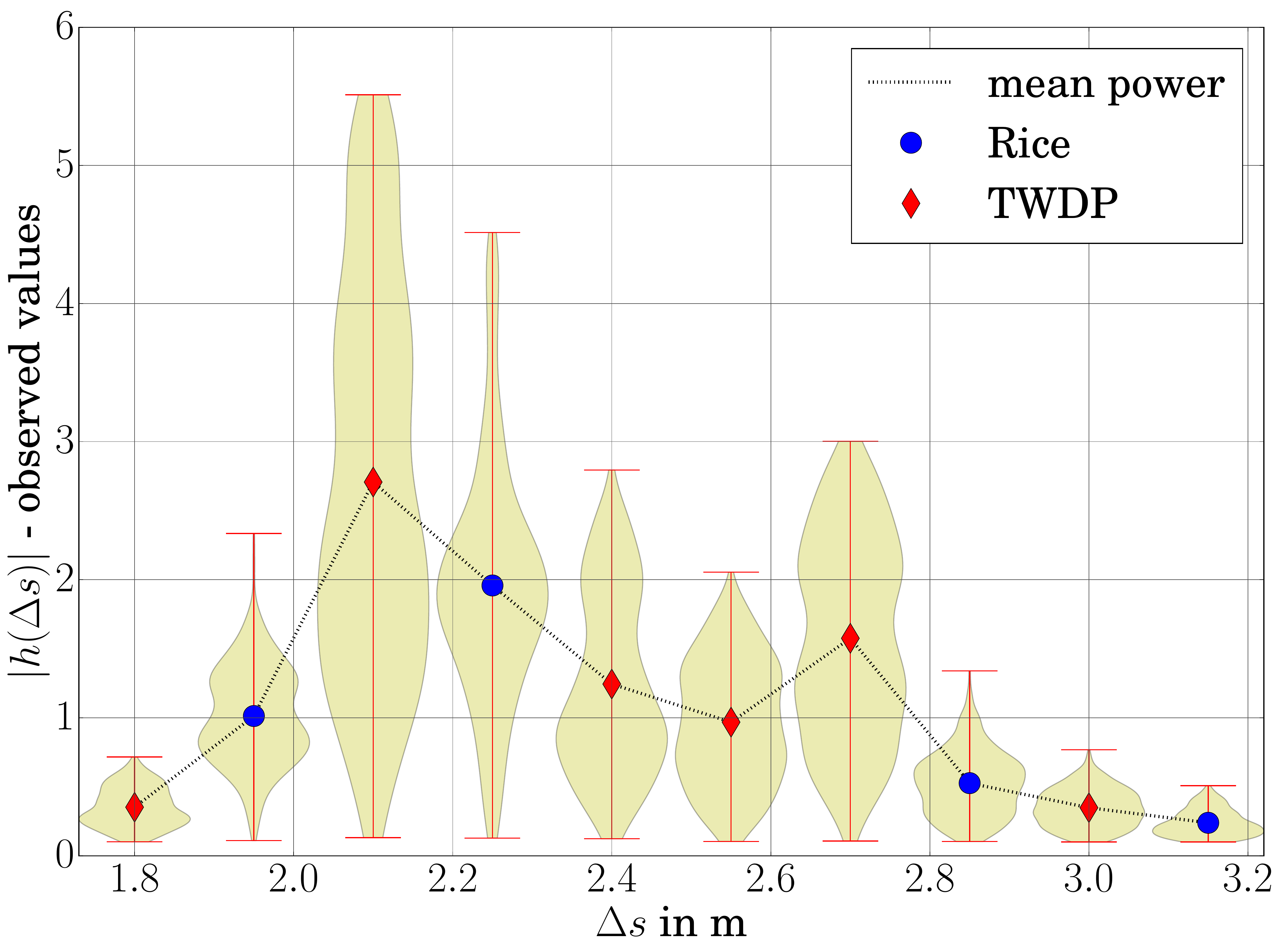}
\caption{{\bf Violin-plot of the CIR time-gated for the wall reflection.}
	This figure shows a zoom-in of the gray highlighted region in Fig.~\ref{fig:CIR}.
	In contrast to Fig.~\ref{fig:CIR}, the y-axis is in linear scale. Thereby the violin plot indicates the distribution at each tap.
    The marker shows the mean value. The marker style codes best fitting distribution.     
     \label{fig:CIRzoom}}
\end{figure}
The gray  highlighted region of Fig.~\ref{fig:CIR} (top part) shows a reflection cluster that corresponds to the excess distance of the wall reflection.
The distributions of each channel tap are represented by a violin plot in Fig.~\ref{fig:CIRzoom}.
A violin plot illustrates the distribution estimated via Gaussian kernels~\cite{hintze1998violin}.
Fig.~\ref{fig:CIRzoom} clearly demonstrates that the \ac{TWDP}-decided distributions have multiple modes. 
The AIC decisions are plotted as markers at the mean power levels.
 
\begin{figure}[H]
\centering
\includegraphics[width=0.65\textwidth]{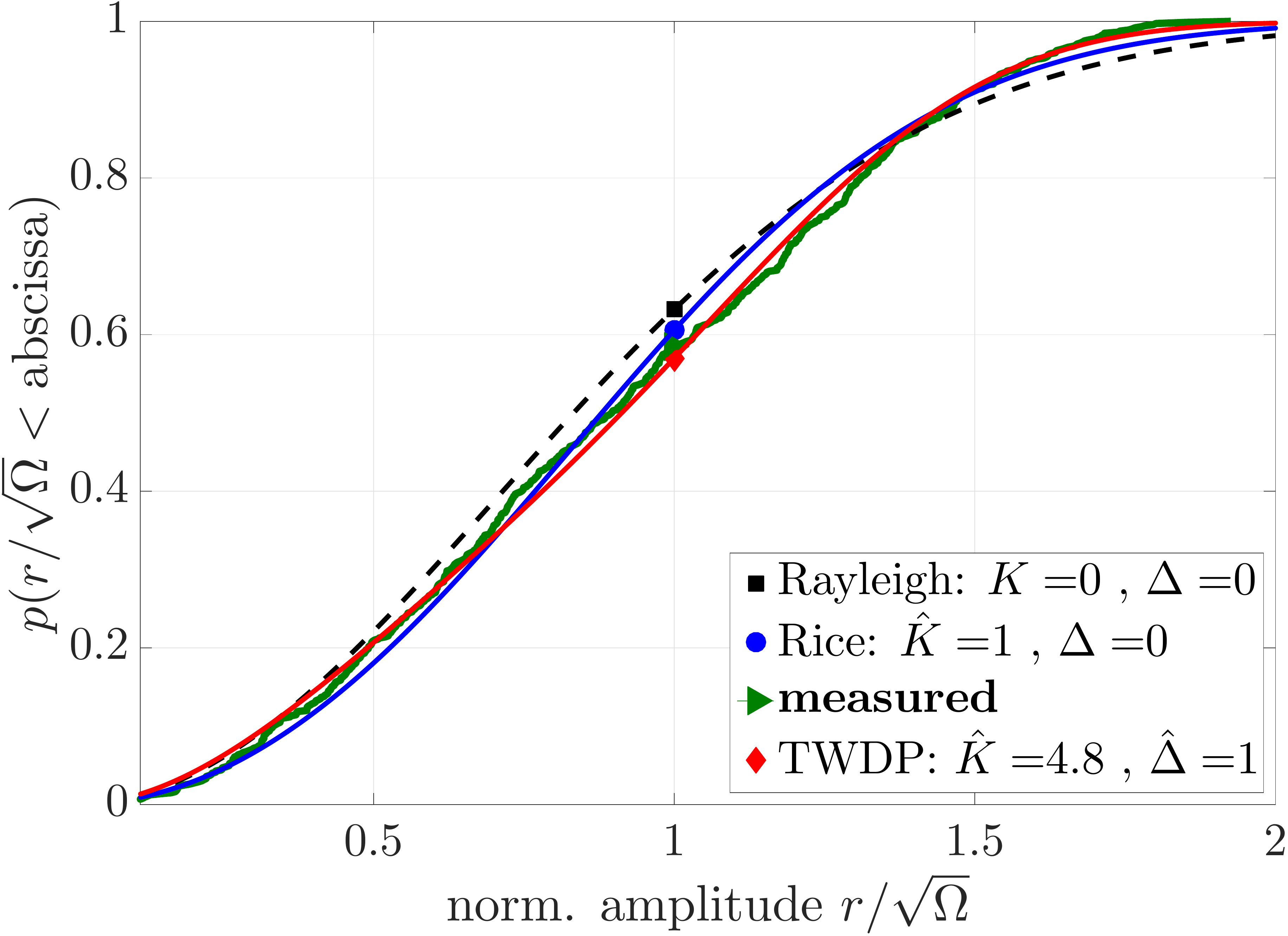}
\caption{{\bf CDF: Distribution fitting for spatial measurement data, time gated by the channel tap at $2.5$\,m, $\varphi=160^\circ$, and $\theta=110^\circ$.}
     Note that, similar to the fitting result in Fig.~\ref{fig:cdf}, the estimated Rician K-factor is again much smaller than the TWDP K-factor.  \label{fig:CIRtap}}
\end{figure}

We evaluated the fading statistic in space for $\varphi=160^\circ$ at the channel tap corresponding to approximately $2.5$\,m excess distance. 
This channel tap is mid in the cluster belonging to the wall reflection.
Fig.~\ref{fig:CIRtap} clearly shows a \ac{TWDP} fading behaviour, confirmed by \ac{AIC}.

\section{Conclusion}

We demonstrate, by means of model selection and hypothesis testing, that \ac{TWDP} fading explains observed indoor millimetre wave channels.
Rician fits of reported studies must be considered with caution.
As two exemplary fits, in Figs.~\ref{fig:cdf} and \ref{fig:CIRtap}, show, Rician K-factors tend to be much smaller than their \ac{TWDP} companions.
There is more power in the specular components than is predicted by the Rician fit.
The \ac{TWDP} fading fit accounts for a possible cancellation of two specular waves.
Our results are verified through two independent measurement campaigns.
For \ac{MC1} and \ac{MC2} we even used different \ac{RF}\,--\,hardware.
While \ac{MC1} was limited to results in the frequency-domain, \ac{MC2} allowed a careful study in the spatial-domain and the time-domain.

Having this strong evidence at hand, we claim that the \ac{TWDP} fading model is more accurate to describe \ac{mmWave} indoor channels.
The flexibility of this model allows furthermore to obtain Rician fading ($\Delta=0$) and Rayleigh fading ($K=0$) results with the same channel model.

In link-level simulations, \ac{TWDP} fading with $\Delta=1$ shows a worse \ac{BER} than Rayleigh fading.
To demonstrate this known effect~\cite{frolik2007case,frolik2008appropriate,frolik2009compact,matolak2011worse,bakir2009diversity}, we provide a \ac{BER} plot in the Appendix.
Rayleigh fading can hence not be used as worst case bound, especially for \ac{mmWave} scenarios.

\section*{Appendix - BER and capacity loss for \ac{TWDP} fading}

\label{sec:link}

\begin{figure}[H]
\includegraphics[width=0.65\textwidth]{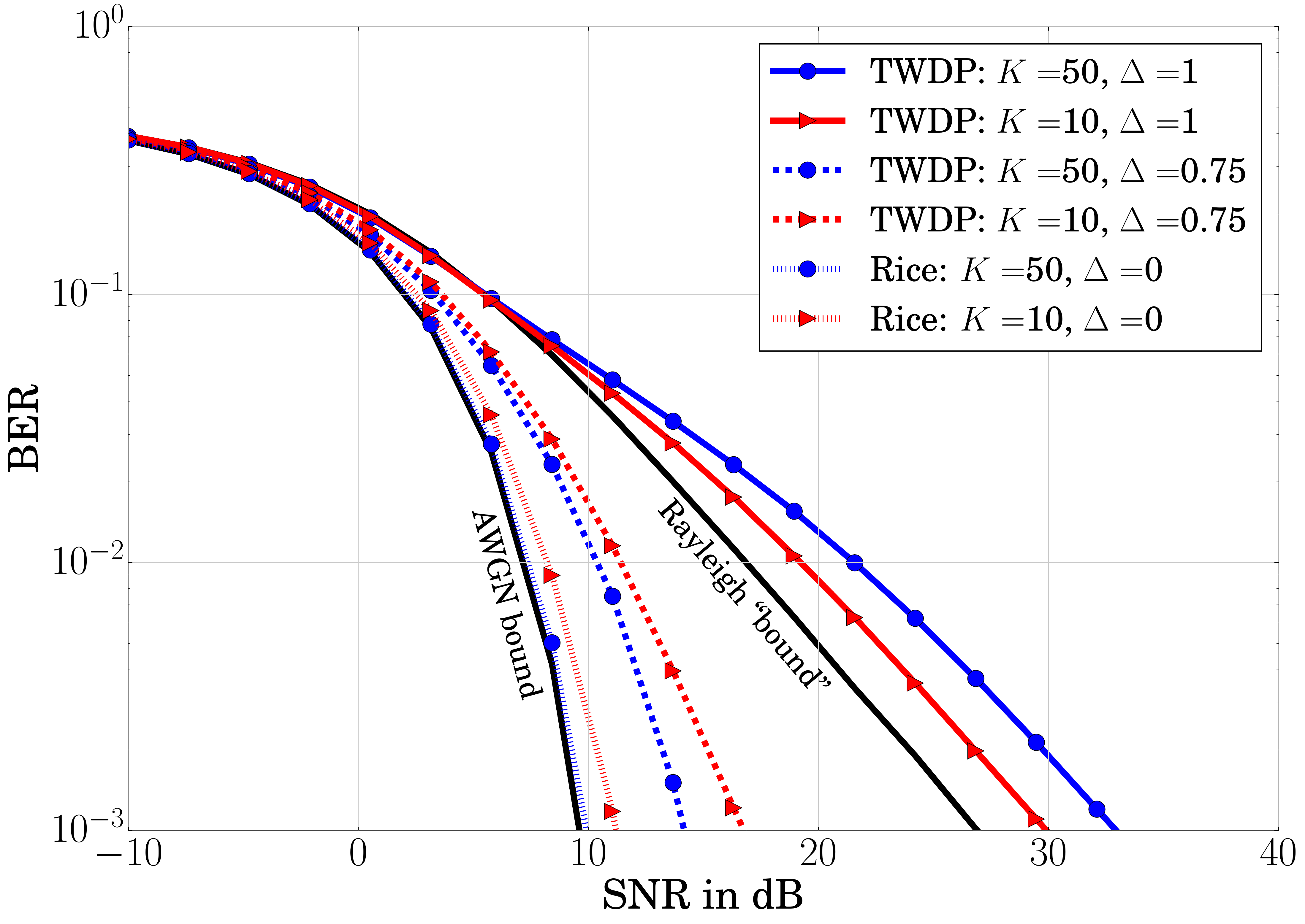}
\caption{{\bf Simulation: Bit error ratio of 4-QAM with {TWDP} fading. }
    The BER performance of TWDP fading potentially lies above the normally chosen ``Rayleigh bound'', for example for $\Delta=1$. 
   \label{fig:BER}}
\end{figure}

We simulate the \ac{BER} of 4-QAM transmissions with Gray-mapping and \ac{TWDP}-fading. 
The channel is perfectly known to the receiver.
The receiver employs a zero-forcing equalizer.
The symbols are normalised to symbol power of one, the SNR is the inverse noise power.
Our simulations assume frequency flat fading and the only channel tap is generated according to a \ac{TWDP} statistic. 
A fading channel tap, given $K$ and $\Delta$, is thus simulated as described by Equation (\ref{eq:TWDPasSum}). 
We simulate \ac{TWDP} flat-fading with independent channel realisations.
Rayleigh fading and Rician fading are included as limiting cases and are simulated as references as well, see Fig.~\ref{fig:BER}.
As $\Delta$ increases, the \ac{BER}-performance gets worse and worse, confer the dashed line for $\Delta=0.5$. 
Finally, Fig.~\ref{fig:BER} shows the worse-than-Rayleigh regions~\cite{matolak2011worse} for $\Delta =1$. 

At this point, we would like to point out that the uncoded BER, of course, has little significance.
The maximum capacity loss $\delta_C$ (in bit/s/Hz) occurs as $K \rightarrow \infty$ and is bounded by~\cite{rao2015mgf}
\be
\delta_C(K \rightarrow \infty,\Delta) = 1 - \log_2 (1+\sqrt{1-\Delta^2}) \le 1~.
\ee

%%%%%%%%%%%%%%%%%%%%%%%%%%%%%%%%%%%%%%%%%%%%%%
%%                                          %%
%% Backmatter begins here                   %%
%%                                          %%
%%%%%%%%%%%%%%%%%%%%%%%%%%%%%%%%%%%%%%%%%%%%%%

\begin{acronym}[mmWave]
\acro{DFT}{discrete Fourier transform}
\acro{mmWave}{millimetre wave}
\acro{AWG}{arbitrary waveform generator}
\acro{PLL}{phase-locked loop}
\acro{SNR}{signal-to-noise ratio}
\acro{SA}{signal analyser}
\acro{LO}{local oscillator}
\acro{RF}{radio frequency}
\acro{RX}{receiver}
\acro{TX}{transmitter}
\acro{CDF}{cumulative distribution function}
\acro{AIC}{Akaike's information criterion}
\acro{TWDP}{two-wave with diffuse power}
\acro{LOS}{line-of-sight}
\acro{NLOS}{non-line-of-sight}
\acro{IQ}{in-phase and quadrature}
\acro{CDF}{cumulative distribution function}
\acro{PDF}{probability density function}
\acro{BER}{bit error ratio}
\acro{MC1}{first measurement campaign}
\acro{MC2}{second measurement campaign}
\acro{VNA}{vector network analyser}
\acro{CIR}{channel impulse response}
\acro{MPC}{multipath component}
\end{acronym}

% if your bibliography is in bibtex format, use those commands:
\bibliographystyle{unsrt} % Style BST file (bmc-mathphys, vancouver, spbasic).
\bibliography{TWDP}      % Bibliography file (usually '*.bib' )
% for author-year bibliography (bmc-mathphys or spbasic)
% a) write to bib file (bmc-mathphys only)
% @settings{label, options="nameyear"}
% b) uncomment next line
%\nocite{label}

\end{document}